\documentclass[prd,onecolumn,a4paper,showkeys]{revtex4}
\usepackage{graphicx}

\begin{document}

\newcommand{\be}{\begin{equation}}
\newcommand{\ee}{\end{equation}}
\newcommand{\bq}{\begin{eqnarray}}
\newcommand{\eq}{\end{eqnarray}}
\newcommand{\bsq}{\begin{subequations}}
\newcommand{\esq}{\end{subequations}}
\newcommand{\bc}{\begin{center}}
\newcommand{\ec}{\end{center}}
\newcommand{\al}{\alpha}

\title{Cosmic String Evolution in Higher Dimensions}

\author{A. Avgoustidis}
\email[Electronic address: ]{A.Avgoustidis@damtp.cam.ac.uk}
\affiliation{Department of Applied Mathematics and Theoretical Physics,
Centre for Mathematical Sciences,\\ University of Cambridge,
Wilberforce Road, Cambridge CB3 0WA, United Kingdom}
\author{E.P.S. Shellard}
\email[Electronic address: ]{E.P.S.Shellard@damtp.cam.ac.uk}
\affiliation{Department of Applied Mathematics and Theoretical Physics,
Centre for Mathematical Sciences,\\ University of Cambridge,
Wilberforce Road, Cambridge CB3 0WA, United Kingdom}

\begin{abstract}
We obtain the equations of motion for cosmic strings in extensions
of the 3+1 FRW model with extra dimensions.  From these we derive  
a generalisation of the Velocity-dependent One-Scale (VOS) model for  
cosmic string network evolution which we apply, first, to a   
higher-dimensional isotropic $D+1$ FRW model and, second, to a   
$3+1$ FRW model with static flat extra dimensions. In the former   
case the string network does not achieve a scaling regime because   
of the diminishing rate of string intersections ($D>3$), but this   
can be avoided in the latter case by considering compact, small   
extra dimensions, for which there is a reduced but still appreciable   
string intercommuting probability. We note that the velocity 
components lying in the three expanding dimensions are Hubble-damped,    
whereas those in the static extra dimensions are only very weakly   
damped. This leads to the pathological possibility, in principle,   
that string motion in the three infinite dimensions can come to   
a halt preventing the strings from intersecting, with the result   
that scaling is not achieved and the strings irreversibly dominate   
the early universe. We note criteria by which this can be avoided,   
notably if the spatial structure of the network becomes essentially   
three-dimensional, as is expected for string networks produced in   
brane inflation. Applying our model to a brane inflation setting, we   
find scaling solutions in which the effective 3D string motion does   
not necessarily stop, but it is slowed down because of the excitations  
trapped in the extra dimensions. These effects are likely to influence   
cosmic string network evolution for a long period after formation and   
we discuss their more general implications.   
\end{abstract}

\preprint{}
\maketitle

\section{\label{intro}Introduction}
 There has been a recent resurgence of interest in cosmic strings  
 both for theoretical and observational reasons. Of particular
 interest is the generic possibility that cosmic strings can be   
 produced at the end of an inflationary phase in models of brane   
 inflation \cite{SarTye,PogTWW}. The evolution of the network of the   
 strings created in these models can be very different from the   
 standard field theory case, thus providing a potential observational  
 window on superstring physics \cite{JoStoTye2,Vilenk,PolchProb}. 
 On the observational side, there are perennial cases of astrophysical 
 phenomena for which cosmic strings have been invoked as an explanation 
 in the absence of some more orthodox mechanism; recent examples are  
 two peculiar gravitational lensing events \cite{Sazhin,Sazhin1,Schild, 
 Kibblerev} but which need further independent follow-up. However,   
 expected improvements in observational data, particularly from high   
 resolution CMB experiments and gravitational wave detectors, present   
 us with the very real prospect of detecting or constraining cosmic   
 strings over a wide range of predicted energy scales (see, for example,   
 \cite{PogTWW, DamVil}). Further theoretical motivation has come from a   
 recent phenomenological study of SUSY GUT models \cite{Jeannerot}, 
 which again found generic cosmic string production (for all cases which 
 solve the monopole problem).

 It is the recent work on cosmic strings appearing in higher dimensional   
 theories, such as brane inflation, which primarily motivates our present  
 study.  For spacetime dimension greater than four, strings no longer   
 generically collide, so that loop production will be highly suppressed. 
 Loops radiate away energy from the long string network, so this 
 suppression will result in a much higher density of cosmic strings   
 than in the usual $3+1$ dimensional case. Jones, Stoica and Tye   
 \cite{JoStoTye2} have estimated this enhancement by using a    
 three-dimensional one-scale model and introducing an intercommuting    
 probability $P<1$ to account for the fact that strings generically   
 miss each other due to the presence of extra dimensions.  They suggest  
 that the enhancement on the energy density of the string network is of  
 order $P^{-2}$, which can be orders of magnitude different than the    
 usual case. 
    
 This approach however does not take into account string velocities 
 in the extra dimensions.  In general, cosmic strings are subject to   
 the constraint that the average velocity squared of string segments   
 must be less than $1/2$. Thus the fact that strings are moving in the   
 extra dimensions will slow down their apparent three-dimensional   
 motion. One might naively expect that velocities in the infinite   
 dimensions will be redshifted by the expansion, while velocities in   
 the compact dimensions will not if these dimensions are static. Hence  
 there is the cosmologically dangerous possibility that velocities in   
 the extra dimensions will accumulate and dominate while string motion    
 in the three infinite dimensions will come to a halt. This can occur   
 before SUSY breaking, when the fields corresponding to the string  
 position in the extra dimensions are expected to become massive. However,  
 if 3D string motion stops for long enough in the early universe, then the  
 strings would irreversibly dominate the energy density of the universe  
 making a subsequent hot big bang model impossible.  Thus we need to   
 go beyond the simple analysis presented to date in order to gain a   
 better quantitative understanding of cosmic string evolution in higher 
 dimensional spacetimes, notably taking into account the important role  
 of velocities in the extra dimensions.   

 String evolution in three spatial dimensions has been studied by various  
 authors. Kibble \cite{Kibble} described string networks by a single  
 lengthscale, the `correlation length', and showed that it evolves  
 towards a scaling solution in which it stays constant with respect  
 to the horizon size. Bennett \cite{Bennett,Bennett1} later  
 modified this `one-scale' model with similar conclusions, subject 
 to a condition on the efficiency of small loop production. The 
 existence and stability of this scaling solution was verified by  
 numerical simulations \cite{BenBouch,AllShel}. These studies also  
 revealed new physics at smaller scales, in particular the accumulation  
 of significant small-scale structure on strings, which results in  
 loop production at much smaller scales than initially thought. 

 To try to incorporate small-scale structure in analytic models, a 
 number of different approaches have been attempted. These include a 
 `kink-counting' model \cite{AllCald,Austin}, a functional approach 
 \cite{Embacher}, a `three-scale' model \cite{AusCopKib} and a  
 `wiggly' model \cite{thesis}. Including small-scale structure  
 in analytical models comes with the cost of introducing several   
 extra parameters, which need to be fixed by simulations.

 However, the large-scale properties of string networks can be 
 quantitatively described by the Velocity-dependent One-Scale (VOS) 
 model \cite{vos0,vos,vosk}, which does not suffer from this problem. 
 By introducing a variable rms string velocity, the VOS model extends    
 its validity from the friction dominated regime at early times,   
 through the matter-radiation transition to $\Lambda$-domination at   
 late times , thus describing the complete cosmological history of   
 string networks. Though it does not directly model small-scale   
 structure, it provides a `thermodynamic' large-scale description 
 of cosmic string evolution, which agrees remarkably well with high   
 resolution numerical simulations.  
 
 The purpose of this paper is to extend the VOS model to spacetimes of 
 higher dimension. Although strong motivation is provided by brane    
 inflation, where the extra dimensions are small and stabilised, we   
 intend to keep the discussion as general as possible to include   
 time-varying extra dimensions, as for example in a $(D+1)$-dimensional   
 FRW universe. 

 The paper is organised as follows. In section \ref{dynamics} we discuss
 cosmic string dynamics in a FRW spacetime with isotropic and flat (but
 possibly expanding) extra dimensions. Starting from the Nambu-Goto
 action in this spacetime we derive the equations of motion as well as an
 expression for the energy of a cosmic string network. In section   
 \ref{evolution} we derive the averaged equations describing the evolution   
 of cosmic strings in that $(D+1)$-dimensional spacetime. After we briefly   
 review the $3+1$ VOS model, we extend it to $D+1$ dimensions and comment    
 on some qualitative features of solutions. In section \ref{braneinflation}
 we consider possible application of this extra-dimensional VOS (EDVOS) model 
 to the case of brane inflation and discuss the dependence of the results   
 on various parameters as for example the intercommuting probability of  
 strings. We conclude in section \ref{conc}. Finally, there is an Appendix   
 where we derive approximate formulae for the momentum parameters of the
 $(D+1)$-dimensional VOS model.

\section{\label{dynamics}Cosmic String Dynamics in Higher Dimensions}

 We consider a cosmic string propagating in a $D+1$ dimensional
 spacetime with metric $g_{\mu\nu}\, \,(\mu,\nu=0,1,2,\ldots,D)$. In 
 the limit that its thickness is much smaller than its radius of
 curvature, the string can be regarded as a one-dimensional object
 with a world history described by a two-dimensional spacetime surface,
 the string worldsheet 
 \be\label{worldsheet}        
  x^\mu=x^\mu(\zeta^\al)\,, \quad\al=0,1 \,. 
 \ee     
 The dynamics is given by the Nambu-Goto action
 \be\label{nambu}   
  S=-\mu \! \int \! \sqrt{-\gamma}\, d^2\zeta
 \ee     
 where $\mu$ is the string tension and $\gamma$ is the determinant of  
 $\gamma_{\alpha\beta}=g_{\mu\nu}\partial_\alpha x^\mu \partial_\beta  
 x^\nu$, the pullback metric on the worldsheet.  
 
 The equations of motion for the fields $x^\mu$ obtained from this
 action are given by
 \be\label{eom}   
  \nabla^2 x^\mu + \Gamma^\mu_{\nu\lambda} \gamma^{\alpha\beta}  
  \partial_\alpha x^\nu \partial_\beta x^\lambda = 0     
 \ee
 where $\Gamma^\mu_{\nu\lambda}$ is the ($D+1$)-dimensional Christoffel  
 symbol
 \be\label{Christoffel}       
  \Gamma^\mu_{\nu\lambda}=\frac{1}{2}g^{\mu\kappa} \left(\partial_\lambda 
  g_{\kappa\nu} + \partial_\nu g_{\kappa\lambda} - \partial_\kappa  
  g_{\nu\lambda} \right)    
 \ee   
 and $\nabla^2 x^\mu$ the covariant Laplacian of the worldsheet fields  
 $x^\mu$ given by    
 \be\label{lapl}   
  \nabla^2 x^\mu = \frac{1}{\sqrt{-\gamma}}\,\partial_\alpha (\sqrt{-\gamma}
  \gamma^{\alpha\beta} \partial_\beta x^\mu) \,.  
 \ee         
 By varying the action with respect to the background metric $g_{\mu\nu}$
 we obtain a spacetime energy-momentum tensor
 \be\label{emt}    
  T^{\mu\nu}=\frac{1}{\sqrt{-g}}\,\mu \! \int \! d^2\zeta 
  \sqrt{-\gamma} \gamma^{\alpha\beta}\partial_\alpha x^\mu 
  \partial_\beta x^\nu \, \delta^{(4)}(x^\lambda-x^\lambda(\zeta^\alpha))\,. 
 \ee    

 We wish to study the evolution of cosmic strings in a FRW universe with
 $D-3$ extra dimensions. For simplicity we choose the following metric,
 allowing toroidal compactification of the extra dimensions 
 \be\label{metric}
  ds^2=N(t)^2 dt^2-a(t)^2 d{\bf x}^2-b(t)^2 d{\bf l}^2
 \ee
 where $t\!\equiv\! x^0$, ${\bf x}\!\equiv\! x^i$ with $i=1,2,3$ and
 $\,{\bf l} \! \equiv\! x^{\ell}$ with $\ell=4,5,\!...,D$. 
 The lapse function $N(t)$ allows us to switch from cosmic ($N(t)\!=
 \!1$) to conformal time by simply setting $N(t)\!=\!a(t)$. For
 stabilised compact extra dimensions, $b(t)$ is set to a constant and the
 coordinates $\bf l$ are periodically identified. Alternatively, the
 scalefactor $b(t)$ of the `$\ell$-space' can be set to $a(t)$ for
 a generalised $(D+1)$-dimensional FRW universe or even more generally 
 be allowed to evolve independently of $a(t)$. 
 
 The action (\ref{nambu}) is invariant under worldsheet  
 reparametrisations, which we can use to choose a gauge. In flat 
 space either the `light-cone' $(\zeta^0=t,\,\zeta^1=z-t)$ or the
 `conformal' gauge $(\dot x\cdot x^{\prime}\!=\!0,\,\dot x^2 \! + \!
 {x^{\prime}}^2\!=\!0)$ is usually chosen but for cosmological
 backgrounds it is convenient to work in the gauge
 \be\label{gauge}
  \zeta^0=t, \quad \dot x\cdot x^{\prime}=0 
 \ee      
 where dots and primes denote derivatives with respect to the timelike
 and spacelike worldsheet coordinates $\zeta^0$ and $\zeta^1$ respectively.
 By choosing this gauge we identify spacetime and worldsheet times, while 
 we impose that the vector $\dot{ x}$ is perpendicular to the string  
 tangent, thus representing the physically observable velocity.   

 In this gauge the equations of motion (\ref{eom}) in the spacetime 
 (\ref{metric}) are
 \bq  
  &&\dot\epsilon=-N^{-2}\epsilon \left\{ N\dot N + a\dot a \left[
    \dot{\bf x}^2 - {\left(\frac{{\bf x}^{\prime}}{\epsilon}\right)}^2  
    \right] + b\dot b \left[\dot{\bf l}^2 - {\left(\frac{{\bf l}^
    {\prime}}{\epsilon}\right)}^2 \right] \right\}
    \label{eombeps}\\     
  &&\ddot {\bf x} + \left\{ \frac{2\dot a}{a}-N^{-2}\left\{ N\dot N+a\dot
    a \left[\dot{\bf x}^2 - {\left(\frac{{\bf x}^{\prime}}{\epsilon}
    \right)}^2\right] + b\dot b \left[\dot{\bf l}^2 - {\left(\frac{{\bf l}^
    {\prime}}{\epsilon}\right)}^2 \right] \right\} \right\} \dot {\bf x}
    ={\left(\frac{{\bf x}^{\prime}}{\epsilon}\right)}^{\prime}
    \epsilon^{-1}
    \label{eombx}\\     
  &&\ddot {\bf l} + \left\{ \frac{2\dot b}{b}-N^{-2}\left\{ N\dot N+a\dot
    a \left[\dot{\bf x}^2 - {\left(\frac{{\bf x}^{\prime}}{\epsilon}
    \right)}^2\right] + b\dot b \left[\dot{\bf l}^2 - {\left(\frac{{\bf l}^
    {\prime}}{\epsilon}\right)}^2 \right] \right\} \right\} \dot {\bf l}
    ={\left(\frac{{\bf l}^{\prime}}{\epsilon}\right)}^{\prime}
    \epsilon^{-1}
    \label{eombl} 
 \eq  
 where $\epsilon$ is a scalar, the energy per unit coordinate length,
 defined by
 \be\label{epsilon}
  \epsilon=\frac{{-x^{\prime}}^2}{\sqrt{-\gamma}}={\left(\frac{a^2{{\bf
  x}^{\prime}}^2+b^2{{\bf l}^{\prime}}^2}{N^2-a^2{\dot{\bf x}}^2-b^2{\dot
  {\bf l}}^2}\right)}^{1/2}\, .  
 \ee
 The energy-momentum tensor (\ref{emt}) for the metric (\ref{metric})
 in the same gauge becomes
 \be\label{emtb}     
  T^{\mu\nu}=\frac{1}{N a^3 b^{D-3}}\,\mu\int \! d\zeta\, (\epsilon \dot
  x^{\mu} \dot x^{\nu} - \epsilon^{-1} {x^\prime}^\mu {x^\prime}^
  \nu) \,\delta^{(D)}\!\left(\,{\bf x} - {\bf x}(\zeta,t)\, , \,\,{\bf l}  
  - {\bf l}(\zeta,t)\,\right)
 \ee
 where we have defined $\zeta\equiv\zeta^1$.        
                     
 The energy of the cosmic string can be defined from $T^{\mu\nu}$ as
 \be\label{Edef}    
  E=\int_{t=\rm{const}} \! \sqrt{h}\, n_\mu n_\nu T^{\mu\nu} d^3
  {\bf x} \, d^{D-3} {\bf l}
 \ee     
 where $n_\mu\!=\!(N,{\bf 0})$ is the normal covector to the spacelike
 $D$-dimensional surface $t=\rm{const}$ and $h$ is the determinant of the
 metric on that surface related to the volume element by
 \be\label{h}
  \frac{1}{D!}\,\sqrt{|g|}\,\epsilon_{\mu\mu_{1}\mu_{2}\ldots\mu_{D}}
  \, n^\mu dx^{\mu_{1}} \wedge dx^{\mu_{2}} \wedge \ldots dx^{\mu_{D}} 
  = \sqrt{h} \, d^3{\bf x} \, d^{D-3} {\bf l}    \,.
 \ee 
 With equation (\ref{emtb}) the energy becomes
 \be\label{E}     
  E(t)=N(t)\, \mu \! \int \! \epsilon\, d\zeta\,.
 \ee     
    
 Note that when ${\bf l}\!=\!{\bf 0}$ and $N(t)\!=\!a(t)$, equations
 (\ref{eombeps}-\ref{eombl}) and (\ref{E}) reduce to the usual
 equations of motion and energy definition for a string in an FRW
 universe written in conformal time \cite{book}. These have been used
 to study cosmic string evolution in an expanding (3+1)-dimensional
 universe. In the next section we discuss cosmic string evolution
 in higher dimensions, based on equations (\ref{eombeps}-\ref{eombl})
 and (\ref{E}).

 \section{\label{evolution}Cosmic String Evolution}

 In this section we discuss the cosmological evolution of strings. 
 After reviewing some basic methods and results in three dimensions, we  
 derive the equations describing the evolution of cosmic strings in 
 a higher dimensional spacetime with metric (\ref{metric}), giving
 particular attention to the case of an isotropic, ($D+1$)-dimensional  
 FRW universe ($b(t)\!=\!a(t)$), and that of stabilised extra dimensions
 $b(t)\!=\!\rm{const}$. Starting from equations (\ref{eombeps}-\ref{eombl})
 and (\ref{E}) we write down a higher dimensional extension of the
 Velocity-dependent One-Scale (VOS) model \cite{vosk}, which has been
 successfully used to provide an analytic `thermodynamical'
 description of the basic properties of evolving cosmic string networks.  
 Application to cosmologically interesting cases is also discussed.

  \subsection{Basics: Evolution in 3+1 Dimensions}

  Monte-Carlo simulations of cosmic string formation after symmetry
  breaking phase transitions suggest that to a good approximation the
  strings have the shapes of random walks at the time of formation.
  Such `Brownian' strings can be described by a characteristic
  length $L$, which determines both the typical radius of curvature
  of strings and the typical distance between nearby string segments
  in the network. On average there is a string segment of length $L$
  in each volume $L^3$ and thus the density of the cosmic string
  network at formation is
  \be\label{rho}
   \rho=\frac{\mu L}{L^3}=\frac{\mu}{L^2} \,. 
  \ee
  A heuristic picture of the evolution of the density of the string
  network can be obtained as follows. Assuming that the strings are
  simply stretched by the cosmological expansion we have $\rho \sim
  a(t)^{-2}$. This decays slower than both the matter and radiation
  energy densities and so such non-interacting strings will soon
  dominate the universe.
  
  This picture changes when the effect of interactions is taken into
  account. As the network evolves, the strings collide or curl back
  on themselves creating small loops, which oscillate and radiatively
  decay. Via these interactions energy is lost from the network so that
  string domination can be avoided. Each string segment travels on
  average a distance $L$ before encountering another nearby segment
  in a volume $L^3$. Assuming relativistic motion and that the produced
  loops have an average size $L$, the corresponding energy loss is
  given by $\dot\rho_{\rm{loops}}\approx L^{-4} \mu L$. The energy loss   
  rate equation becomes
  \be\label{rholoss}
   \dot\rho\approx -2\frac{\dot a}{a}\rho - \frac{\rho}{L}\,.
  \ee
  Cosmic string networks are known to evolve towards a `scaling' regime
  in which the characteristic length $L$ stays constant relative to the 
  the horizon $d_H\sim t$ \cite{Kibble}. To see this we set $L=\gamma(t)
  t$ and substitute (\ref{rho}) into (\ref{rholoss}) to obtain
  \be\label{gdot}
   \frac{\dot\gamma}{\gamma}=\frac{1}{2t}\left(2(\beta-1)+\frac{1}{\gamma}
   \right) \,.  
  \ee
  The parameter $\beta$ is related to the scalefactor $a(t)$ by $a(t)   
  \propto t^\beta$ and is equal to $1/2$ and $2/3$ in the radiation  
  and matter eras respectively. Equation (\ref{gdot}) has a scaling
  solution   
  \be\label{slnosm}   
   \gamma=[2(1-\beta)]^{-1}  
  \ee    
  (which depends on cosmology through the expansion exponent $\beta$)     
  demonstrating that the characteristic length scales at a value   
  $L\sim t$. If we start with a high density of strings, intercommuting
  will produce loops reducing the energy of the network, whereas if the   
  initial density is low then there will not be enough intercommuting   
  and $\gamma$ will decrease. Given enough time, the two competing   
  effects of stretching and fragmentation will always reach a   
  steady-state and the scaling regime will be approached. 

  Equation (\ref{gdot}) was derived on physical grounds and it only 
  captures the basic processes involved in string evolution, namely  
  the stretching and intercommuting of strings. It does not take into  
  account other effects like the redshifting of string velocities due 
  to Hubble expansion. However we can derive a more accurate 
  evolution equation for the string energy density based on the  
  Nambu-Goto action. By differentiating the energy (\ref{E}) with 
  respect to time for the case of a 3+1 FRW universe and setting
  $E \propto \rho a^3$ we find 
  \be\label{rhodot3d}
   \dot\rho=-\left(2\frac{\dot a}{a} + 2N^{-2}a\dot a \langle
   \dot{\bf x}^2 \rangle\right)\rho 
  \ee
  where we have defined
  \be
    \left\langle \dot{\bf x}^2 \right\rangle\equiv \frac{\int\!   
    \dot{\bf x}^2 \epsilon d\zeta}{\int\!\epsilon d\zeta}\,.
  \ee
  We also introduce a phenomenological term to account for energy  
  loss through loop production 
  \be\label{phen}
   \dot\rho_{\rm{loops}}=-\frac{\tilde c v\rho}{L}
  \ee
  where $\tilde c$ is the loop production parameter, related to  
  the integral of an appropriate loop production function over all 
  relevant loop sizes (see equation (\ref{phen3dim}) and the discussion  
  above it), and $v$ the average velocity of intercommuting string  
  segments.
 
  Using (\ref{rho}) and setting $L=\gamma(t)\,t$ as before, we 
  obtain the following equation, expressed in physical time t
  \be\label{gdtvos}
   \gamma^{-1}\,\frac{d\gamma}{dt}=\frac{1}{2t}\left(2\beta(1+v^2)-
   2+\frac{\tilde c v}{\gamma}\right)\, .   
  \ee
  Here we defined $v$, the average velocity of string segments, by
  \be\label{v}
   v^2=\left\langle{\frac{d{\bf x}}{d\tau}}^2\right\rangle\equiv 
   \frac{\int\!{\frac{d{\bf x}}{d\tau}}^2\epsilon d\zeta}{\int\!
   \epsilon d\zeta}
  \ee
  with $\tau$ the conformal time $(N(\tau)=a(\tau))$.

  Equation (\ref{gdtvos}) is of the same form as (\ref{gdot}) but
  has an extra correction term $\beta v^2$ accounting for redshifting
  of velocities due to cosmological expansion. It also includes the 
  parameter $\tilde c$, the value of which can be extracted from   
  numerical simulations and it is of order unity \cite{book,vos}.

  To solve (\ref{gdtvos}) we also need an evolution equation for $v$.
  This can be obtained by differentiating (\ref{v}) with respect to
  $\tau$ and using the three-dimensional version of the equation of
  motion for $\bf x$ (\ref{eombx}). The result (expressed in physical
  time $t$) is \cite{vos0}
  \be\label{vdtvos}
   \frac{dv}{dt}=(1-v^2)\left(\frac{k}{R}-2Hv\right)    
  \ee
  where $H$ is the Hubble parameter $\dot a/a \!=\! \beta t^{-1}$,  
  $R$ the average radius of curvature of strings in the network and $k$   
  the momentum parameter (first introduced in \cite{vos0}) defined by  
  \be\label{kvos}
   \frac{kv(1-v^2)}{R}=\left\langle \dot{\bf x}\cdot{\bf u}
   (1-\dot{\bf x}) \right\rangle  
  \ee
  ${\bf u}$ being the curvature vector defined by
  \be\label{curvvos}
   \frac{d^2\bf x}{a\,ds^2}={\bf u}=\frac{1}{R} \hat{\bf u} 
  \ee
  with $ds=\sqrt{{{\bf x}^{\prime}}^2}d\zeta$. For a Brownian network 
  (and within the VOS assumptions) the average radius of curvature $R$   
  is equal to the correlation length $L\equiv\gamma t$. 

  Equation (\ref{vdtvos}) has a clear physical meaning: velocities of
  string segments are produced by string curvature and damped by the
  cosmological expansion. Together with (\ref{gdtvos}) it constitutes
  the Velocity-dependent One-Scale (VOS) model, which has been
  demonstrated to be in very good agreement with numerical simulations
  \cite{vostests}. It has the scaling solution  
  \be\label{solnvos}
    \gamma^2=\frac{k(k+\tilde c)}{4\beta(1-\beta)}  \;\;\;\;\;\;\;\;\;\;
     v^2=\frac{k(1-\beta)}{\beta(k+\tilde c)}
  \ee
  in terms of the expansion exponent $\beta$, the loop production  
  parameter $\tilde c$ and the momentum parameter $k$. 

  The momentum parameter is a measure of the angle between the  
  curvature vector and the velocity of string segments and thus is 
  related to the smoothness of the strings. Slowly moving strings are
  smooth so the velocity is more or less parallel to the curvature vector, 
  resulting in a value of $k$ of order unity. As $v$ increases towards 
  relativistic values the accumulation of small-scale structure renders 
  the strings wiggly. Velocities become uncorrelated to curvature and   
  $k$ decreases. In particular it can be shown analytically that for 
  flat space, where $v^2=1/2$, the momentum parameter vanishes for a wide   
  range of known solutions \cite{vos,thesis}. 

  An accurate ansatz for the momentum parameter $k$ has been proposed
  in \cite{vosk}   
  \be\label{kans3d}
   k = k(v) = \frac{2\sqrt{2}}{\pi} \frac{1-8 v^6}{1+8 v^6} 
  \ee 
  satisfying $k(1/\sqrt{2})=0$. 

  Note that the fact that $v=1/\sqrt{2}$ in flat spacetime, can be 
  shown analytically for closed loops only, but for long strings it  
  is observed in numerical simulations \cite{book}. For expanding    
  or contracting spacetimes, $v$ is less or greater than $1/\sqrt{2}$ 
  respectively. Hence for an expanding universe, string velocities 
  are subject to the constraint
  \be\label{vconstr}
   v^2 \le \frac{1}{2}\,.  
  \ee     
  This fact will be important for our discussion of extra dimensions.

  \subsection{Evolution in Higher Dimensions}
 
   We now proceed to derive macroscopic evolution equations for 
   string networks in higher dimensions, based on equations 
   (\ref{eombeps}-\ref{eombl}) and (\ref{E}), derived in section
   \ref{dynamics}. The result is the extra-dimensional velocity-dependent  
   one-scale (EDVOS) model. 
 
   \subsubsection{\label{evltng}Evolution Equation for $\gamma$} 
   
   We start by differentiating the energy equation (\ref{E}) with
   respect to time and using the equation of motion (\ref{eombeps}) 
   for $\epsilon$. This gives  
   \be\label{Edot}
    \dot E=-\frac{1}{N(t)^2} \left[ a\dot a \left( \langle \dot{\bf x}^2
    \rangle E - \mu N(t)\int\epsilon^{-1} {{\bf x}^{\prime}}^2 d\zeta
    \right) + b\dot b \left( \langle \dot{\bf l}^2 \rangle E - \mu N(t)
    \int\epsilon^{-1} {{\bf l}^{\prime}}^2 d\zeta \right) \right]  
   \ee  
   where we have defined the average of a function $f$ along the string as
   \be\label{average}
    \left\langle f \right\rangle\equiv \frac{\int\! \epsilon f d\zeta}  
    {\int\!\epsilon d\zeta}\,.
   \ee
   We then use $E\propto \rho a^3 b^{D-3}$ and definition (\ref{epsilon})  
   for $\epsilon$ to obtain an evolution equation for the energy density of     
   the string network
   \bq 
    -\frac{\dot\rho}{\rho} = \frac{\dot a}{a} \left[2 + 2N^{-2}a^2 \langle
    \dot{\bf x}^2 \rangle + N^{-2}b^2 \langle \dot{\bf l}^2 \rangle + N^{-2}  
    \left\langle \frac{ {b^2{\bf l}^{\prime}}^2 }{a^2 {{\bf x}^{\prime}}^2 + 
    b^2{{\bf l}^{\prime}}^2 } \left(N^2-a^2 \dot{\bf x}^2 - b^2 \dot{\bf l}^2   
    \right) \right\rangle  \right] \nonumber \\  
    + \frac{\dot b}{b} \left[(D-4) + 2N^{-2}b^2 \langle \dot{\bf l}^2 \rangle  
    + N^{-2}a^2 \langle \dot{\bf x}^2 \rangle + N^{-2} \left\langle \frac{   
    {a^2{\bf x}^{\prime}}^2 }{a^2 {{\bf x}^{\prime}}^2 +  
    b^2{{\bf l}^{\prime}}^2 } \left(N^2-a^2 \dot{\bf x}^2 - b^2 \dot{\bf l}^2
    \right) \right\rangle \right]\, .  \label{rhodot}   
   \eq 
   We would normally proceed by assuming that the string network at formation  
   is Brownian with a characteristic length $L$ and writing  
   \be\label{rhoex}
    \rho=\frac{\mu L}{L^D}=\frac{\mu}{L^{D-1}} \,.  
   \ee
   Unfortunately, since the metric we are considering is not in general   
   isotropic, we do not expect this Brownian structure to be preserved by   
   the evolution. The amount by which string segments are stretched will   
   depend on their orientation and as time passes, the random walk shape 
   of strings will be distorted by the anisotropic expansion. However, there  
   are two interesting special cases in which the strings can remain   
   Brownian. First, the case of isotropic expansion ($a(t)\!=\!b(t)$),    
   corresponding to a generalised $(D+1)$-dimensional FRW universe, and  
   second, a situation where the formation of the string network is   
   localised on an isotropic slice. The latter is the case, for example,   
   in brane inflation \cite{DvalTye,DvalShafSolg,BMNQRZ,Garc-Bell,JoStoTye1}  
   where string formation takes place essentially on a brane and there is an   
   effective three-dimensional description of the evolution. We will consider  
   both cases below.    
 
\medskip
\noindent {\it (i) Isotropic case:} 
   Starting with the isotropic case, we set $b(t)\!=\!a(t)$ in equation  
   (\ref{rhodot}) to obtain an evolution equation for the energy density  
   of a non-interacting string network in a ($D+1$)-dimensional FRW universe  
   \be\label{rhodotFRW}
    \dot\rho=-\frac{\dot a}{a} \left[ (D-1) + 2N^{-2}a^2 \langle
    \dot{\bf x}^2 \rangle + 2 N^{-2}a^2 \langle \dot{\bf l}^2 \rangle    
    \right] \rho \,.   
   \ee
   As in the usual VOS model we can treat string interactions by introducing   
   a phenomenological loop production term. In the three-dimensional case a  
   string segment of size $L$, the correlation length, is expected to travel  
   a distance $L$ before encountering another segment and interacting with it  
   in a volume $L^{-3}$. If the probability of such an interaction producing  
   a loop of length between $\ell$ and $\ell+d \ell$ is given by the scale   
   invariant function $f(\ell/L)$ then the energy loss due to loop production  
   is \cite{book} 
   \be\label{phen3dim}
    \dot\rho_{\rm{loops}}=-\frac{\mu v}{L^3} \int_{0}^{\infty}  
    \frac{d \ell}{L}f(\ell/L) = -\frac{v\rho}{L} \int_{0}^{\infty}  
    \frac{d \ell}{L}f(\ell/L) \equiv -\frac{\tilde c v\rho}{L} \,. 
   \ee 
   When the number of spatial dimensions $D>3$, 
   the string segments will generically not interact after moving a  
   distance $L$. In particular, if the strings have a thickness or    
   capture radius $\delta$ (quantifying how close they need to be in    
   order to interact), then the probability of interacting after moving 
   a distance $L$ (in time $\delta t$) in a volume $L^D$ is   
   \be\label{PinterDdims}   
    \frac{v \delta t}{L} \frac{1}{L^D} \left(\frac{\delta}{L}\right)^{D-3}\,.  
   \ee
   Note that this suppression is essentially the intercommuting 
   probability $P$ which we shall further discuss later, though here it 
   is time dependent. 
   The loop production term is then  
   \be\label{phenDdim}
    \dot\rho_{\rm{loops}}=-\frac{\mu v}{L^D} \left(\frac{\delta}{L}  
    \right)^{D-3} \int_{0}^{\infty}\frac{d \ell}{L}f(\ell/L) =      
    -\frac{\tilde c v\rho}{L} \left(\frac{\delta}{L}\right)^{D-3}  
    \equiv -\frac{v\rho}{L} {\tilde c}_D      
   \ee   
   where we have defined  
   \be\label{c_D} 
    {\tilde c}_D=\tilde c \left(\frac{\delta}{L}\right)^{D-3}\,.    
   \ee     
   Including this loop production term in equation (\ref{rhodotFRW}) and  
   setting as in the three-dimensional case $L=\gamma(t)t$, where $t$ is   
   the physical time ($N(t)=1$) we obtain   
   \be\label{gdtFRW}
    \gamma^{-1}\,\frac{d\gamma}{dt}=\frac{1}{(D-1)\,t} \left[\beta \left(
    (D-1)+2v^2 \right) + \frac{v \tilde c_D}{\gamma} - (D-1) \right]\,. 
   \ee    
   Here $v^2={v_x}^2+{v_\ell}^2$ where $v_x$ and $v_\ell$, the rms peculiar  
   velocities of string segments in the $x$ and the $\ell$ directions 
   respectively, are defined by   
   \be\label{vxvl}
    v_x^2=\left\langle\frac{a^2}{N^2} \dot{\bf x}^2\right\rangle
    \;\;\;\; {\rm{and}} \;\;\;\;\; v_\ell^2=\left\langle\frac{b^2}{N^2}  
    \dot{\bf l}^2 \right\rangle=\left\langle\frac{a^2}{N^2}\dot{\bf l}^2  
    \right\rangle   
   \ee  
   since $b(t)\!=\!a(t)$ in the isotropic case.  

   Note that since $\tilde c_D$ is proportional to $L^{-(D-3)} \!=\!  
   (\gamma t)^{-(D-3)}$, the loop production term $v \tilde c_D/\gamma$  
   is explicitly time dependent and no scaling solution $\gamma\!=\!  
   \rm{const}$ exists. In particular, as time increases the loop   
   production term becomes smaller and smaller, reflecting the fact    
   that strings cannot find each other and interact in more than three    
   spatial dimensions. String evolution in this isotropic $D+1$-dimensional   
   case will be discussed further in section \ref{solutions}. 
  
\medskip
\noindent {\it (ii) Anisotropic case:}
   We now consider the case were $b(t)\!\ne\! a(t)$ in (\ref{metric}) but   
   the string network is produced on a FRW slice (the $x$-space) of the   
   spacetime. If the extra dimensions are larger than the correlation length    
   and the strings are allowed to move in the bulk after formation, then   
   the anisotropic expansion will distort their Brownian structure and 
   they will be more difficult to characterise analytically. We leave
   this case for subsequent discussion (see the final section). The   
   situation is more straightforward if the extra dimensions are  
   compactified at a size smaller than the correlation length which we  
   shall assume from now on. We can then have an effective   
   three-dimensional description in which the strings will remain   
   Brownian due to isotropy in the three-dimensional FRW space. The   
   presence of the extra dimensions will reduce the intercommuting  
   probability in the effective three-dimensional description, but 
   we shall assume it is non-zero. In principle, we can also have    
   time-varying extra dimensions, as long as their size remains much    
   smaller than the correlation length at all times. However, for    
   simplicity we only consider static extra dimensions and set $b(t)\!=\!1$,    
   assuming stabilisation by some unspecified mechanism.

   For such cases we can substitute (\ref{rho}) in equation (\ref{rhodot})  
   to obtain an effective three-dimensional evolution equation for the 
   energy density, which however takes into account velocities in all 
   $D$ spatial dimensions. As before we also introduce a loop production   
   term $-\frac{\tilde c v_x \rho}{L}$ and set $L\equiv\gamma(t)\, t$   
   where $t$ is the physical time ($N(t)=1$). We thus obtain an effective    
   evolution equation for $\gamma(t)$  
   \be\label{gdt}    
    \gamma^{-1}\,\frac{d\gamma}{dt}=\frac{1}{2\,t} \left\{\beta \left[
    \left(2+w_\ell^2 \right)+\left(2-w_\ell^2 \right){v_x}^2+\left(1-  
    w_\ell^2 \right){v_\ell}^2 \right] -2 + \frac{\tilde c v_x}{\gamma}   
    \right\}   \,.
   \ee   
   In deriving this expression, we have assumed    
   \be\label{approx}   
    \left\langle \frac{ {{\bf l}^{\prime}}^2 }{a^2  
    {{\bf x}^{\prime}}^2 + {{\bf l}^{\prime}}^2 } \left(N^2-a^2
    \dot{\bf x}^2 - \dot{\bf l}^2 \right) \right\rangle \simeq
    \left\langle \frac{ {{\bf l}^{\prime}}^2 }{a^2
    {{\bf x}^{\prime}}^2 + {{\bf l}^{\prime}}^2 } \right\rangle
    \left\langle \left(N^2-a^2 \dot{\bf x}^2 - \dot
    {\bf l}^2 \right) \right\rangle   
   \ee
   and we have defined an orientation parameter 
   \be\label{wl}
    w_\ell=\sqrt{ \left\langle \frac{ {{\bf l}^{\prime}}^2 }    
    {a^2 {{\bf x}^{\prime}}^2 + {{\bf l}^{\prime}}^2 }\right\rangle }
   \ee  
   which quantifies the degree to which the strings lie in the extra  
   $\ell$ dimensions.  
    
   Note the differences between (\ref{gdt}) and the analogous equation  
   for a 3+1 FRW universe (\ref{gdtvos}). Comparing (\ref{gdt}) to 
   (\ref{gdtvos}) we see that the coefficients of dilution and 
   $x$-velocity redshift are modified by $w_\ell^2$, while there  
   is an additional term accounting for the redshifting of velocities in  
   the static $\ell$-space. We interpret this term as arising from    
   the conservation of momentum: an expanding string segment which is   
   transverse to the $\ell$-space and moves in the $\ell$ directions   
   is slowed down because its effective mass increases as the string
   is stretched in the $x$-direction.

   Another important difference with the three-dimensional case is that the  
   loop production parameter $\tilde c$ is now suppressed by a factor  
   of order $(\delta/R_\ell)^{(D-3)}$, where $R_\ell$ is the compactification  
   scale. If $R_\ell$ was greater than the correlation length $L$, then the  
   suppression factor would instead be $(\delta/L)^{(D-3)}$, which is 
   time-dependent. Thus there would be no scaling solution initially, as   
   in the $(D+1)$-dimensional FRW case we discussed above (see later   
   discussion for the asymptotic evolution in this case). Here however,  
   we assumed the compactification scale is much less than the correlation   
   length and so the suppression factor can be treated as a constant    
   intercommuting probability. For a string-theoretic calculation of   
   this probability see Ref.~\cite{PolchProb}.

   \subsubsection{\label{evltnvxvl}Evolution Equations for $v_x$ and $v_\ell$}
   
   As in the three-dimensional case we also need to know how the average   
   velocities of string segments evolve in time.
   We begin by considering the case of static extra dimensions $b(t)=1$.  
   Working in conformal time $\tau$ ($N(\tau)=a(\tau)$),
   we start by differentiating ${v_x}^2\!=\!\frac{\int\! \dot{\bf x}^2  
   \epsilon d\zeta}{\int\!\epsilon d\zeta}$ with respect to $\tau$ and 
   using the equation of motion (\ref{eombx}) for the fields $\bf x$. 
   We obtain    
   \be\label{v_xdot}   
    \begin{array}{c}   
    v_x \dot v_x=\frac{1}{2} \left[ 2\frac{\dot a}{a}\left(
    {\langle\dot {\bf x}^2 \rangle}^2 - \langle\dot {\bf x}^4 
    \rangle\right) + \frac{\dot a}{a^3}\left( {\langle \dot
    {\bf x}^2 \rangle}^2 {\langle \dot {\bf l}^2 \rangle}^2 -
    \langle \dot{\bf x}^2 \dot {\bf l}^2 \rangle \right) +
    \frac{\dot a}{a^3} \left\langle\left(\langle\dot {\bf x}^2
    \rangle - \dot {\bf x}^2 \right){{\bf l}^{\prime}}^2\epsilon^
    {-2}\right\rangle \right] \\
    +a\left\langle \dot {\bf x}\cdot  {\bf u}\left(
    1-\dot{\bf x}^2-\frac{\dot{\bf l}^2}{a^2}\right) \right\rangle
    -2\frac{\dot a}{a} \langle\dot {\bf x}^2\rangle + 2\frac{\dot a}  
    {a}\langle\dot {\bf x}^4\rangle + \frac{\dot a}{a^3}\langle\dot
    {\bf l}^2 \dot {\bf x}^2 \rangle \\
    + \frac{\dot a}{a} \left\langle \frac{ {{\bf l}^{\prime}}^2 }
    { a^2 {{\bf x}^{\prime}}^2 + {{\bf l}^{\prime}}^2 } \left(1- 
    \dot{\bf x}^2 - \frac{\dot{\bf l}^2}{a^2} \right) {\dot{\bf x}}^2 
    \right \rangle - \left\langle \dot{\bf x}\cdot{\bf x}^{\prime}  
    \frac{ \dot{\bf x}\cdot\dot{\bf x}^{\prime}+\dot{\bf l} \cdot  
    \dot{\bf l}^{\prime}/a^2 }{ {{\bf x}^{\prime}}^2 + {{\bf l}^
    {\prime}}^2/a^2 } \right\rangle     
    \end{array} 
   \ee 
   where ${\bf u}$ is the curvature vector defined by   
   \be\label{curv}     
    \frac{d^2 (a{\bf x},{\bf l})}{a^2 ds^2} = {\bf u} = \frac{1}{R}
    \hat{\bf u}
   \ee     
   with $ds=\sqrt{\left({{\bf x}^{\prime}}^2+{{\bf l}^{\prime}}^2/a^2
   \right)}d\zeta$ and $R$ the radius of curvature, equal to the 
   correlation length $L\equiv\gamma t$ for a VOS model Brownian network.   
   
   The first three terms in  equation (\ref{v_xdot}) (first line) 
   can be neglected because their only effect is to slightly modify   
   the coefficients of all the remaining terms in the expression (but    
   for the last). Numerical confirmation of the smallness of such terms   
   (in particular the first term in the case of conventional FRW) was  
   presented in \cite{vos}. The remaining terms (again except the last  
   one) are of the same form as in the (3+1)-dimensional FRW case but   
   now they also include terms dependent on the velocity in the    
   $\ell$-space and an extra term which gives corrections of order    
   $\left\langle \frac{ {{\bf l}^{\prime}}^2 } { a^2 {{\bf x}^{\prime}}^2   
   + {{\bf l}^{\prime}}^2 } \right\rangle$.

   The last term arises from the artificial splitting of the $D$-dimensional  
   velocity into an $x$ and an $\ell$ part. It is absent in the usual  
   (3+1)-dimensional case because of the gauge condition $\dot{\bf x}  
   \cdot{\bf x}^{\prime}=0$. Here the gauge condition is $\dot{\bf x}  
   \cdot{\bf x}^{\prime}-\dot {\bf l}\cdot{\bf l}^{\prime}/a^2=0$ and   
   thus $\dot{\bf x}\cdot {\bf x}^{\prime}$ does not identically vanish.   
   However, we expect its value along the string to change sign randomly   
   with no large-scale correlations. Hence, averaged over the whole string   
   network, $\dot{\bf x}\cdot{\bf x}^{\prime}$ is expected to vanish so   
   that we can neglect the last term of (\ref{v_xdot}). This was tested    
   numerically for a three-dimensional FRW model, splitting $v$ into a    
   $xy$- and a $z$-part and neglecting the corresponding terms in the    
   $xy$- and $z$-velocity equations. The evolution of the system using    
   these equations was found numerically to be very close to the    
   corresponding evolution using the full three-dimensional velocity    
   equation, with the scaling values of $v$ and $\gamma$ agreeing to   
   three significant figures.
   
   Again using the approximation (\ref{approx}) and switching to 
   physical time $t$, we can write (\ref{v_xdot}) in a much more     
   elegant and useful form, 
   \be\label{v_xdt}
    v_x \frac{d v_x}{dt}=\frac{k_x v_x}{R}(1-v^2)-\left(2-w_\ell^2 
    \right) H{v_x}^2 (1-v^2)-H{v_x}^2{v_\ell}^2 
   \ee
   where $w_\ell$ is given in (\ref{wl}) and $k_x$ is defined by    
   \be\label{k_x}
    \frac{k_x v_x (1-v^2)}{R}=\left\langle \dot {\bf x}\cdot  {\bf u}
    \left( 1-\dot{\bf x}^2-\frac{\dot{\bf l}^2}{a^2}\right) \right\rangle\,.
   \ee     
   Similarly the evolution equation for $v_\ell$ reads
   \be\label{v_ldt}
    v_\ell \frac{d v_\ell}{dt}=\frac{k_\ell v_\ell}{R}(1-v^2)-\left(1- 
    w_\ell^2 \right)H{v_\ell}^2(1-v^2)+H{v_\ell}^2{v_x}^2
   \ee
   with $k_\ell$ defined in analogy to $k_x$ 
   \be\label{k_l}
    \frac{k_\ell v_\ell (1-v^2)}{R}=\left\langle \frac{\dot{\bf l}}{a}
    \cdot  {\bf u} \left( 1-\dot{\bf x}^2-\frac{\dot{\bf l}^2}{a^2}  
    \right) \right\rangle\,.
   \ee    
   We see that, as may have been anticipated from (\ref{rhodot}), the  
   $v_x$ damping term is very different from that of $v_\ell$. To   
   demonstrate this we neglect $w_\ell^2$ corrections and note    
   that $H{v_x}^2$ in (\ref{v_xdt}) comes with a factor of    
   $(2-2{v_x}^2-{v_\ell}^2)$ but $H{v_\ell}^2$ in (\ref{v_ldt}) has    
   a factor of only $(1-2{v_x}^2-{v_\ell}^2)$ (which can cancel almost   
   to zero). This result will be important for the discussion in the    
   next section. 

   We can also write down an evolution equation for the total
   velocity $v$, by differentiating $v^2=\langle(\dot{\bf x}^2,
   \dot{\bf l}^2/a^2)\rangle$. The result is simply the sum of  
   (\ref{v_xdt}) and (\ref{v_ldt}) with   
   \be\label{k}   
    k_x v_x + k_\ell v_\ell = k v
   \ee
   where $k$ has a similar definition to (\ref{k_x}) and
   (\ref{k_l}) involving the dot product of $(\dot{\bf x},
   \dot{\bf l}/a^2)$ with the curvature vector $\bf u$. Equation  
   (\ref{k}) follows from the linearity of the dot product and the 
   integral.    

   The momentum parameter $k$, as in the (3+1)-dimensional case 
   measures the angle between the curvature vector and the velocity
   of string segments, thus providing a measure of the smoothness of
   the strings. For smooth strings (the non-relativistic limit) the 
   velocity of string segments is expected to be more or less 
   parallel to the curvature vector $\bf u$ corresponding to $k$ of 
   order unity. For relativistic velocities however, small-scale  
   structure accumulates and the strings become wiggly. The parameter   
   $k$ is then expected to approach zero.      

   Similarly, $k_x$ and $k_l$ provide a measure of the  
   angle between the curvature vector and the $x$-velocity or the  
   $\ell$-velocity respectively. Therefore they encode two effects:
   the wiggliness of the strings and the extent to which the curvature
   vector $\bf u$ lies in the $x$ or $\ell$ subspace. 
 
   Following \cite{vosk} we split ${\bf v}\equiv(\dot{\bf x},\dot{\bf l}  
   /a)$ into a `curvature' component ${\bf v}_c$ produced during the 
   last correlation time and a `left-over' component ${\bf v}_p$, coming
   from previous accelerations. We can then derive the following
   approximate formulae for $k, k_x$ and $k_\ell$, valid in the  
   relativistic regime (see Appendix \ref{appendix})  
   \bq
    && k\simeq \frac{1-8 v^6} { \left(1+8 v^6 \right)^{1/2} 
       \left(1+8(D-2)v^6 \right)^{1/2} }
      \label{kans} \\  
    && k_x\simeq \frac{v_{xc}}{v_c} \frac{1-8 v^6} { \left(
       1+8 v^6 \right)^{1/2} \left(1+8(D-2)v^6 \right)^{1/2} }  
      \label{k_xans} \\   
    && k_\ell\simeq \frac{v_{\ell c}} {v_c} \frac{1-8 v^6} { \left(
       1+8 v^6 \right)^{1/2} \left(1+8(D-2)v^6 \right)^{1/2} } 
      \label{k_lans}\,.    
   \eq       
   Note that for flat spacetime, the condition $v^2\!=\!1/2$ still holds 
   and so $k(1/\sqrt{2})\!=\!0$. For an expanding universe the velocities 
   of string segments are subject to the constraint
   \be\label{vxvlconstr}
    v^2 = v_x^2+v_\ell^2 \le \frac{1}{2}\,.  
   \ee

   Finally, we note that the velocity evolution equation in the case of   
   an isotropic ($D+1$)-dimensional FRW universe ($b(t)\!=\!a(t)$) can  
   be similarly found    
   \be\label{vdtFRW}
    \frac{dv}{dt}=(1-v^2)\left(\frac{k}{R}-2Hv\right)\,.  
   \ee
   Note that this equation does not depend on $w_\ell$. However, in  
   the case $b(t)\!=\!1$ (eqns (\ref{v_xdt}), (\ref{v_ldt})) there is an  
   explicit dependence on $w_\ell$, so we need to know how it evolves.   
   Unfortunately, the search for an evolution equation for $w_\ell$ is   
   problematic as we discuss below.   

   \subsubsection{\label{probwl}The $w_\ell$ equation and higher
                  order terms}   

   Equations (\ref{gdt}), (\ref{v_xdt}) and(\ref{v_ldt}) depend on the   
   orientation parameter $w_\ell$. We interpret this as the degree   
   to which the strings lie in the extra dimensions, that is, a hidden   
   small-scale structure parameter. To make a fully closed system  
   of equations we also need an evolution equation for $w_\ell$.  
   
   In analogy to our treatment of string velocities, we can try to 
   obtain an evolution equation for $w_\ell$ by differentiating 
   its definition (\ref{wl}) with respect to $\tau$. Unfortunately   
   this produces terms of the form
   \be\label{badterms}   
    \frac{\dot a}{a}\left( {\langle {{\bf x}^\prime}^2 \rangle} {\langle 
    \dot {\bf x}^2 \rangle} - \langle {{\bf x}^\prime}^2 \dot {\bf x}^2 
    \rangle \right), \;\; \frac{\dot a}{a}\left( {\langle {{\bf x}^\prime}^2
    \rangle}^2  - \langle {{\bf x}^\prime}^4 \rangle \right), \;\; 
    \frac{\dot a}{a}\left( {\langle\dot {\bf x}^2 \rangle} {\langle  
    {{\bf l}^\prime}^2 / a^2 \rangle} - \langle \dot{\bf x}^2   
    {{\bf l}^\prime}^2 /a^2 \rangle \right), \;\;  \rm{etc}. 
   \ee  
   These purely statistical terms are higher order and cannot be   
   easily determined. We found terms of similar kind in the    
   $x$-velocity equation (\ref{v_xdot}), but in that case their   
   only effect was to `renormalise' the coefficients of other terms   
   in the equation. Here, however, there are no such terms to be   
   renormalised and the small differences between these undetermined    
   statistical terms contribute at leading order.     

   However, there are special cases in which an evolution equation for 
   $w_\ell$ is not needed. For example, the evolution of a string 
   network in a higher dimensional FRW universe is described by equations 
   (\ref{gdtFRW}) and (\ref{vdtFRW}), which do not depend on $w_\ell$.   
   In particular, isotropy suggests that string segments would have equal  
   probability of moving in any direction and one would expect $w_\ell$  
   to be a constant, namely the square root of the ratio of the number of   
   extra dimensions, $D-3$, over the total number of spatial dimensions $D$.  

   For the case of small $w_\ell$ we note that since it appears in  
   our equations only through factors of $(1-w_\ell^2)$ and 
   $(2-w_\ell^2)$ modifying various coefficients, its evolution in 
   time can be ignored as long as it stays small at all times. This is  
   conceivable in cases of FRW universes with static extra dimensions,   
   where the Hubble expansion of the three FRW dimensions will tend to    
   reduce the value of $w_\ell$. 
  
   Finally, in cases that none of the above are applicable, it may be  
   possible to find an ansatz for $w_\ell$ as a function of the 
   different velocity components, based on physical arguments. We   
   shall consider all three possibilities in the following discussion.      
   
   \subsubsection{\label{solutions}General Features of Solutions}

\begin{figure}
    \includegraphics[height=2.7in,width=3in]{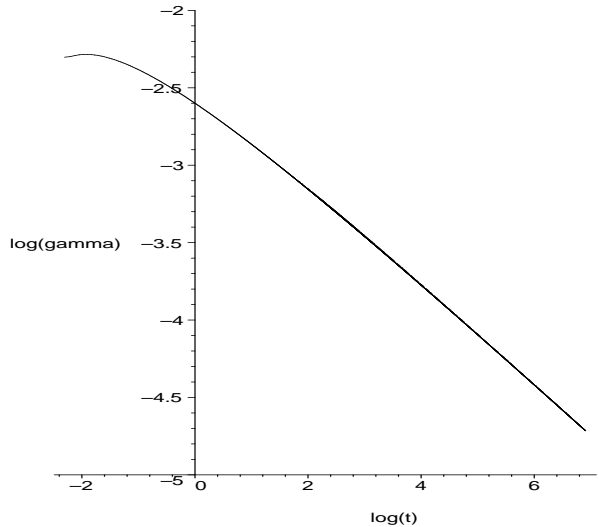}
    \caption{\label{gFRW} Evolution of $\gamma$ in a log-log plot for   
             a string network evolving in a $(4+1)$-dimensional   
             FRW universe. The capture radius $\delta$ of the strings  
             has been assumed to be equal to the initial correlation  
             length and the exponent parameter $\beta$ has been set to  
             1/2. There is no scaling solution and at late times   
             $log(\gamma)$ has a constant slope of $-1/3$, in agreement   
             with our asymptotic solution (\ref{asymptsol}).}     
   \end{figure}

\noindent {\it (i) Isotropic case:}
   We begin by considering the isotropic case $b(t)\!=\!a(t)$ 
   corresponding to a $(D+1)$-dimensional FRW universe. The energy  
   density evolution is described by equation (\ref{gdtFRW}).  
   Comparing this to the corresponding equation in 3+1 dimensions 
   (\ref{gdtvos}) we see that the effect of extra FRW dimensions  
   is to modify two of the terms in (\ref{gdtvos}) by a factor of 
   $2/(D-1)$. This reflects the fact that in three spatial dimensions, 
   the density of the network is inversely proportional to the square  
   of the correlation length $L$ (equation (\ref{rho})), whereas in  
   $D>3$ dimensions it is inversely proportional to the $(D-1)^{\rm{th}}$   
   power of $L$ (equation (\ref{rhoex})). The energy density in the    
   higher dimensional case decreases faster, as there are more   
   expanding dimensions (also see equation (\ref{rhodotFRW})). Of   
   course there is a second, more dramatic, effect related to string   
   interactions. As noted in \ref{evltng}, the loop production parameter  
   $\tilde c_D$ in the higher dimensional case is much smaller, since it    
   is harder for two strings to find each other and intercommute in $D>4$.  
   In fact, $\tilde c_D$ is proportional to $(\gamma t)^{-(D-3)}$ and as  
   a result equation (\ref{gdtFRW}) does not have a scaling solution   
   of the form $\gamma=\rm{const}$. In particular, the loop production  
   term becomes smaller as time increases and $\gamma$ will always 
   decrease, so that strings will dominate the energy density of the  
   universe. This can be seen in Fig. \ref{gFRW} were equations  
   (\ref{gdtFRW}) and (\ref{vdtFRW}) with $\tilde c_D$ given by   
   (\ref{c_D}) have been solved numerically for $D=4$. 

   The asymptotic solution for $L$ in a $D$-dimensional isotropic model can  
   be easily found to be 
   \be\label{asymptsol} 
    \frac{L}{L_o} = \left(\frac {t}{t_o}\right) ^{\beta D/(D-1)} =    
    \left(\frac {a}{a_o}\right) ^{D/(D-1)}\, , \qquad v = \frac{1}{\sqrt{2}} 
    \, . 
   \ee    
   Contrast this EDVOS model result with simple conformal stretching of the   
   string network $L\propto a$, obtained by assuming $v=0$. In the EDVOS model  
   if we start with $v\ll 1$ we still find $L\propto a$, but we also discover  
   that $v$ increases according to $v \propto t^{1-\beta}$. Thus, conformal  
   stretching is only a transient solution, which will be followed by one   
   with non-negligible $v$. The corresponding scaling result is   
   \be\label{asymptsolv}
    L \propto t^{\beta\frac{D+2v^2-1}{D-1}} \propto a^{\frac{D+2v^2-1}{D-1}}\,  
    , \qquad v ={\rm const} \, .
   \ee   
   In flat space $v^2$ cannot exceed $1/2$ but in an expanding space this   
   upper value is somewhat reduced. Since the horizon is expanding linearly   
   in time, the correlation length quickly falls behind and the cosmological 
   expansion becomes irrelevant so that $v^2\rightarrow 1/2$. Equation   
   (\ref{asymptsolv}) then implies (\ref{asymptsol}) asymptotically.  
   
   Note that (\ref{asymptsolv}) means that the string energy density    
   scales as $\rho_s \propto L^{-(D-1)}\propto a^{-D+1-2v^2}$, which decays  
   slower than radiation $\rho_{\rm rad}\propto a^{-(D+1)}$. Therefore, if   
   this regime persists for long enough, the strings will eventually dominate   
   the universe. The Friedmann equation for a string dominated $D+1$ FRW  
   universe yields  
   \be\label{scalefact} 
    a \propto t^{2/(D+2v^2-1)}\equiv t^{\beta}\, .  
   \ee
   Substituting in (\ref{asymptsolv}) we finally obtain   
   \be\label{Loft} 
    L \propto t^{2/(D-1)}\, .
   \ee   
   The above reproduces and generalises the recent findings of Ref.   
   \cite{MartNonInt} for $D=3$. In particular, we recover non-intercommuting  
   strings in the linear scaling solution $L\propto t$ (i.e. setting $P=0$  
   and $D=3$). 

   In a situation were the extra dimensions are compactified, but expanding  
   isotropically with scalefactor $a(t)$, the solution (\ref{asymptsol})  
   implies that the correlation length will eventually catch up with the   
   size of the extra dimension $R_\ell$ and so the string network will   
   become effectively three-dimensional with ${\tilde c}_D=\tilde c \left(  
   \delta/R_\ell\right)^{D-3}$ where $L>R_\ell \propto a$. Unlike   
   the case of static compact dimensions, which we will discuss, this   
   change is insufficient to prevent string domination over ordinary matter  
   and radiation.  

\noindent {\it (ii) Anisotropic case:} We now turn to the case $b(t)\!=\!1$.  
   The relevant equations are 
   (\ref{gdt}) and (\ref{v_xdt}-\ref{v_ldt}) with $k_x$ and $k_l$   
   approximately given by (\ref{k_xans}-\ref{k_lans}). To demonstrate 
   the general features of solutions, we can keep $w_\ell$ as a 
   \emph{constant} parameter (in the range $0\le w_\ell \le 1$),  
   whose only effect is to modify the coefficients of damping terms   
   in (\ref{gdt}), (\ref{v_xdt}) and (\ref{v_ldt}). We will later 
   consider a simplified version of the EDVOS model, where 
   $w_\ell$ is not constant, but is given by a physically   
   motivated ansatz, which is a function of $v_{\ell}$ and $v_x$.  
  
\begin{figure}  
    \includegraphics[width=3in,keepaspectratio]{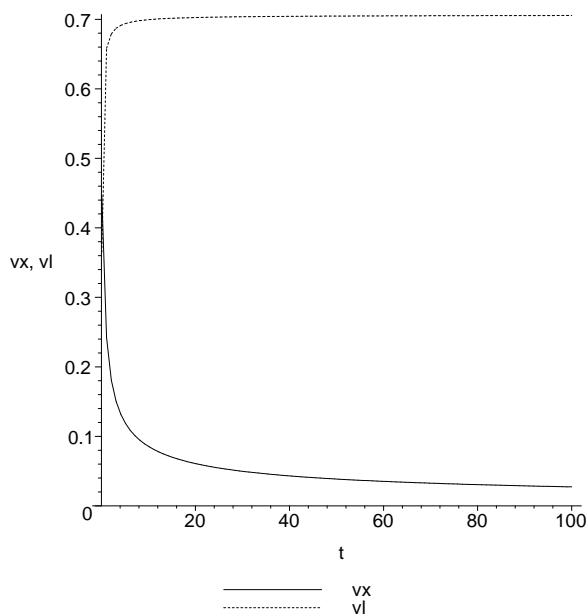}
    \caption{\label{Dcurve} Velocity evolution for a Brownian network  
             in $D$ spatial dimensions, three of which are expanding.   
             The expansion strongly redshifts the velocities in    
             the expanding dimensions ($v_x$), resulting in domination    
             of the extra dimensional velocities ($v_\ell$). The strings   
             will stop moving in the expanding dimensions.}
   \end{figure}

   If the string network at formation is Brownian in $D$ dimensions,    
   then the curvature vector $\bf u$ explores the 
   $x$-space as well as the $\ell$-space and $k_x, k_\ell$ are   
   comparable. Equations (\ref{v_xdt}) for $v_x$ and (\ref{v_ldt})  
   for $v_\ell$ have comparable source terms but $v_\ell$ has a   
   much weaker damping term. Therefore, we expect $v_\ell$ to become   
   much greater than $v_x$ and eventually to drive $v_x$ to zero,   
   because of the constraint ${v_x}^2+{v_\ell}^2\le 1/2$. We conclude  
   that expansion of the $x$-space together with the fact that  
   the $\ell$-space is not expanding will halt string motion in 
   the $x$-space. This can be verified numerically (Fig. \ref{Dcurve}) 
   by solving (\ref{v_xdt}),(\ref{v_ldt}) with the approximations   
   $\frac{v_{xc}}{v_c}\simeq \frac{v_x}{v}$ and   
   $\frac{v_{\ell c}}{v_c}\simeq \frac{v_\ell}{v}$ in   
   (\ref{k_xans}-\ref{k_lans}). These are valid approximations 
   in both the non-relativistic and relativistic limits, as long 
   as the curvature vector explores all $D$ spatial dimensions.  
   Thus the strings in this case will soon become non-interacting  
   and, if this regime were to last long enough, dominate the universe  
   irreversibly. However, there are two important caveats in this case,   
   namely that anisotropic expansion would soon distort the simple  
   Brownian network structure and that the intercommuting probability   
   for a string network in $D$ spatial dimensions would be small and    
   decreasing (as discussed above). For compact extra dimensions, this   
   regime will end when the correlation length becomes larger than the   
   compact dimension $L> R_\ell$, after which point the evolution becomes  
   effectively three-dimensional (we will return to this point later).

\begin{figure} 
    \includegraphics[width=3in,keepaspectratio]{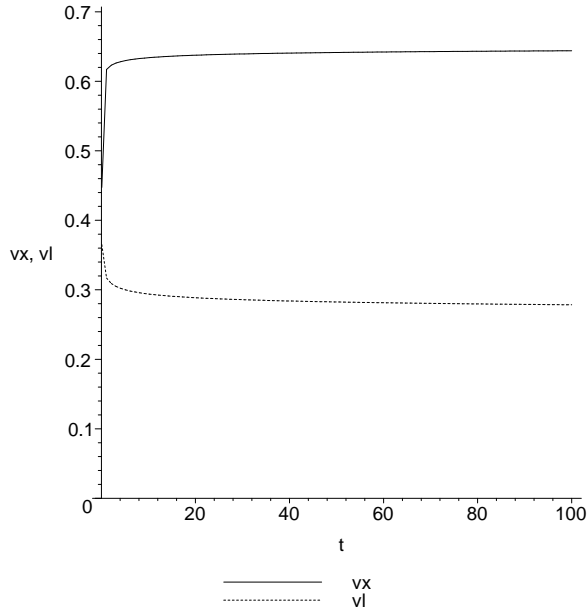}
    \caption{\label{3curve} Velocity evolution for a Brownian network     
             formed on a FRW slice. The curvature vector lies in the   
             space of the three expanding dimensions, so there is no  
             source for the extra dimensional velocities $v_\ell$. They  
             are given by their initial condition, redshifted weakly by  
             the expansion. The three-dimensional velocity $v_x$ will not  
             vanish but it can be significantly smaller than in the   
             purely three-dimensional case $v\simeq 0.7$.}
   \end{figure}

   The situation is different if the curvature vector effectively  
   lies in the $x$-space, as may be the case in brane inflation 
   (see next section) where the formation of the string network  
   is constrained on an effectively $(3+1)$-dimensional FRW slice.   
   In this case the dot product ${\bf u}\cdot \dot{\bf l}$ is   
   negligible and thus $k_\ell$ is much smaller than $k_x$. Equation  
   (\ref{v_ldt}) has essentially no source term. We expect to see   
   $v_x>v_\ell$ with $v_\ell$ given by its initial condition, weakly   
   damped according to equation (\ref{v_ldt}). This was verified 
   numerically (Fig. \ref{3curve}) by setting $\frac{v_{xc}}{v_c}  
   \simeq \frac{v_x}{v}$, as before, and $\frac{v_{\ell c}}{v_c}
   \ll 1$. The effect of increasing $\frac{v_{\ell c}}{v_c}$ was 
   also considered. It was found that there is a critical value  
   of $\frac{v_{\ell c}}{v_c}\simeq 0.15$ for which the scaling  
   values of $v_x$ and $v_\ell$ become equal. Above this critical 
   value $v_\ell$ eventually dominates and we return to the previous   
   regime, with $k_\ell$ and $k_x$ comparable. In particular, for   
   $\frac{v_{\ell c}}{v_c}\simeq 0.5$, $v_\ell$ soon reaches a    
   `relativistic' value, driving $v_x$ to zero. It may seem that    
   if the strings are free to explore the extra dimensions after    
   formation we may run into the same problems we had before,     
   namely a practically zero intercommuting probability and the  
   loss of the Brownian structure of the string network. However,   
   the extra dimensions can be small and compact, in which case the    
   probability of intercommuting can still be appreciable and the    
   network is to a good approximation Brownian in three dimensions.      

   Finally we consider the effect of varying the parameter
   $w_\ell$. We find that $\gamma$ is insensitive to changes  
   in $w_\ell^2$ between $0$ and $0.2$ but is reduced by   
   approximately $20\%$ for $w_\ell^2=0.5$. On the other hand   
   $v_{\ell}$ depends on $w_\ell$ more strongly. In particular,  
   changing $w_\ell^2$ from $0$ to $0.01$ increases the scaling  
   value of $v_\ell$ by approximately $0.5\%$ but a further   
   increase of $w_\ell^2$ to $0.1$ increases $v_\ell$ typically by  
   $30\%$ (this does not strongly affect $v_x$, which is approximately  
   equal to $\sqrt{1/2-{v_\ell}^2}$). As a result, the effective   
   three-dimensional behaviour of the evolving network is insensitive    
   to the value of $w_\ell^2$, at least when it is smaller than 0.1    
   or so. There is a critical value of about 0.2 above which the sign    
   of the damping term in equation (\ref{v_ldt}) becomes positive and  
   $v_\ell$ begins to increase. This apparently unphysical effect    
   signifies the fact that the constant $w_\ell$ approximation    
   breaks down for relatively large values of $w_\ell$. In the   
   following we discuss a simplified version of the extra-dimensional     
   VOS model with $w_\ell$ given by a function of $v_{\ell}$ and    
   $v_x$, where this unphysical behaviour does not occur.     

   \subsubsection{\label{simplemodel}A Simplified Model}
   
   We can obtain a simplified version of equations (\ref{gdt}),  
   (\ref{v_xdt}) and (\ref{v_ldt}) by expressing $w_\ell$ as a function  
   of $v_{\ell}$ and $v_x$. String gradients in one direction are produced  
   by velocities in the same direction so one would expect the network to  
   achieve a sort of equipartition between velocities and gradients in the  
   extra dimensions. This motivates the ansatz     
   \be\label{wlsans}
    w_\ell^2=\frac{{v_\ell}^2}{v^2}  
   \ee 
   which mathematically corresponds to    
   \be        
    \left\langle \frac{ \dot{\bf l}^2 } {a^2 \dot{\bf x}^2 + \dot{\bf l}^2   
     }\right\rangle = \left\langle \frac{ {{\bf l}^{\prime}}^2 }
    {a^2 {{\bf x}^{\prime}}^2 + {{\bf l}^{\prime}}^2 }\right\rangle . 
   \ee    
   With this substitution, equations (\ref{gdt}) and (\ref{v_xdt}-\ref{v_ldt}) 
   become
   \bq       
    && \gamma^{-1}\,\frac{d\gamma}{dt}=\frac{1}{2\,t} \left[\beta \left(  
        2+2{v_x}^2+\frac{{v_\ell}^2}{v^2} \right) -2 + \frac{\tilde c v_x}   
        {\gamma}\right]  
       \label{gdtsimple}\\ 
    && v_x \frac{d v_x}{dt}=\frac{1}{t}\left[\frac{k_x v_x}{\gamma} 
        (1-v^2) - \left(2-2{v_x}^2-\frac{{v_\ell}^2}{v^2}\right)  
        \beta{v_x}^2 \right]   
       \label{v_xdtsimple}\\ 
    && v_\ell \frac{d v_\ell}{dt}=\frac{1}{t}\left[\frac{k_\ell v_\ell}  
        {\gamma}(1-v^2)-\left(1-\frac{{v_\ell}^2}{v^2}\right)(1-2v^2)  
        \beta{v_\ell}^2 \right]\,.       
       \label{v_ldtsimple}
   \eq 
   These three equations  
   describe the macroscopic evolution of a Brownian network of   
   strings, produced in a three-dimensional FRW slice, which are then free  
   to explore the extra dimensions. In order for the Brownian structure  
   to be maintained (so that equation (\ref{gdtsimple}) is valid) the  
   extra dimensions need to be compactified at a size smaller than the  
   initial correlation length at the time of formation.   
 
   Note that the unphysical behaviour leading to the change of sign of    
   the damping terms in the $v_\ell$ evolution equation (which is a result  
   of the fact that the constant $w_\ell$ approximation is not valid   
   for large $w_\ell$) is now absent from equation (\ref{v_ldtsimple}).   
   The only way the damping terms can change sign is if $v^2$ exceeds $1/2$,  
   which is not allowed by the constraint (\ref{vxvlconstr}). Similarly   
   the damping terms in equation (\ref{v_xdtsimple}) cannot change sign  
   either.     

   We can now solve (\ref{gdtsimple}-\ref{v_ldtsimple}) together   
   with (\ref{k_xans}-\ref{k_lans}) for $k_x$ and $k_\ell$. Since we are
   considering  the case where the structure of the network at formation is  
   essentially three-dimensional, the curvature vector lies   
   mainly in the $x$-space, so that $k_\ell \ll k_x$ (equivalently   
   $\frac{v_{\ell c}}{v_c}\ll 1$ in equation (\ref{k_lans})). The   
   solutions are qualitatively the same as the corresponding constant   
   $w_\ell$ solutions mentioned above. This is because    
   $w_\ell^2={v_\ell}^2/v^2$ is small (and decreasing with time) so   
   its effect in equations (\ref{gdtsimple}-\ref{v_ldtsimple}) is only   
   to slightly modify coefficients of order unity. We will discuss     
   these solutions in more detail in the next section where we will   
   apply the model to brane inflation.

\section{\label{braneinflation}Application to Brane Inflation}  

 Recently there has been much activity in trying to derive models  
 of cosmological inflation from string theory \cite{DvalTye,  
 DvalShafSolg,BMNQRZ,Garc-Bell,JoStoTye1,KKLMMT,HsuKalPro,
 FirTye,HsuKal,BurClStQue,DeWKaVer,IizTriv,Racetr}, in which there   
 are a large number of scalar fields that can potentially serve as  
 inflatons. One of the most interesting scenarios (especially from 
 the point of view of cosmic string production) is brane inflation  
 \cite{DvalTye,DvalShafSolg,BMNQRZ,Garc-Bell,JoStoTye1}, where the  
 role of the inflaton field is played by the distance between two branes 
 or a brane and an anti-brane. The branes are initially relatively  
 displaced and move towards each other due to an attractive potential 
 arising from the exchange of bulk NS-NS and R-R modes. As they come 
 closer together, open string modes stretching between them start   
 contributing more strongly to the inflaton potential. At a critical  
 distance of order the inverse superstring scale ${M_{\rm{s}}}^{-1}$  
 such modes become tachyonic and inflation ends in a hybrid-inflation-type
 exit. 

 Brane-stacks carry Chan-Paton gauge groups and hence the tachyonic
 instability appearing at brane collision corresponds to a symmetry
 breaking process. This allows the formation of topological defects 
 as lower dimensional branes if the vacuum manifold is non-trivial 
 \cite{Majumdar}. For two brane-stacks of $N$ branes each, the vacuum 
 manifold is isomorphic to $U(N)$, whose only non-trivial homotopy  
 groups are the odd ones, $\pi_{2k-1}$. Thus the topologically  
 allowed defects have \emph{even} codimension $2k$.  

 We need to make sure that cosmologically dangerous defects like  
 monopoles or domain walls are not produced. This is fortunately  
 the case (but see \cite{Matsuda}) as can be seen by the following 
 argument given in Ref.~\cite{SarTye}. In order to describe our    
 (3+1)-dimensional universe the branes must have three infinite    
 spatial dimensions. They can either be D$3$-branes or D$p$-branes    
 with $(p-3)$ dimensions compactified at a size smaller than the    
 horizon \cite{DvalTye}. Thus the Kibble mechanism can only take  
 place in the three infinite dimensions so the codimension of the     
 defects must lie in these dimensions (hence it can only be 1,2  
 or 3). Since the codimension is even, the defects that are produced
 have codimension 2, that is they are D$(p-2)$-branes wrapping the   
 same compact space as the original D$p$-branes. These objects will  
 appear as cosmic strings to a three-dimensional observer.    

 Alternative string production mechanisms (bulk preheating, 
 resonant string formation) have been discussed in \cite{Vilenk},  
 resulting in both D-string and F-string networks, or even an 
 interacting (F-D)-string network (also see \cite{PolchRevis}).    
 String stability was studied in \cite{PolchStab}. Various possible   
 cosmologically relevant metastable strings were found in many    
 models, including $(p,q)$ strings that is, bound states of $p$    
 F-strings and $q$ D-strings. For string stability also see    
 \cite{LebTye}.

 Given these possibilities, cosmic string production can be  
 considered as a generic prediction of brane inflation (though
 models with no cosmic strings have been constructed \cite{Quevedo}). 
 It is then natural to ask what the evolution of the string  
 network would be in these models and whether it is distinctly   
 different from the standard cosmic string evolution, thus providing 
 a potential observational window into superstring physics.  
 We consider this question below.

 \subsection{\label{setup}Basics}    
   
  We have seen that in the context of brane inflation, D-strings 
  are in general D$(p-2)$-branes wrapping the same $(p-3)$-dimensional  
  compactified space as the initial D$p$-branes. Their production
  is localized on the plane of brane collision, but later they are  
  free to explore the $(9-p)$ compact dimensions transverse to the brane. 

  For two strings to collide their worldsheets must intersect. In  
  the 3+1 case the number of spacetime dimensions equals the sum  
  of the dimensions of the two worldsheets and hence strings will
  generally collide \cite{BrandVaf}. For a higher dimensional 
  spacetime, however, this is no longer the case so strings will  
  generally miss. One then expects to be able to model cosmic string  
  evolution in the context of brane inflation by simply introducing 
  an intercommuting probability $P$ in the usual $(3+1)$-dimensional  
  evolution equations. 

  This was done in \cite{JoStoTye2} where an intercommuting 
  probability $P$ ($\lambda/\lambda_0$ in the original paper) was  
  introduced in equation (\ref{gdot}). The new scaling solution
  is $\gamma \simeq P$ (cf $\gamma\sim 1$ in three dimensions),   
  corresponding to an enhancement in the scaling energy density    
  of strings by a factor of $P^{-2}$. This can be orders of    
  magnitude different than the usual 3D case \cite{PolchProb}.  

  There are at least two effects which could significantly 
  alter the above result. The first is related to the
  small-scale structure of the strings. If they are wiggly,  
  they may have more than one opportunities to intercommute   
  during each crossing time. One expects that a probability of   
  $1/3$ or so should have no impact in the scaling of $\gamma$,   
  and thus, contrary to what is sometimes suggested in the   
  literature, strings with such large $P$ could not be   
  observationally distinguished from standard field theory   
  strings. On the other hand, for $P\ll 1$ the scaling value  
  of $\gamma$ should depend more strongly on $P$. The dependence   
  of $\gamma$ on $P$, taking into account small-scale wiggles   
  can only be modelled numerically. Simulations in flat space    
  suggest a scaling $\gamma\simeq \sqrt{P}$ in the range $0.05<P<0.3$    
  \cite{Sak,VilSak,Vilenk}. To deal with this uncertainty we will    
  introduce an {\it effective} intercommuting probability    
  $P_{\rm eff}=f(P)$ as a multiplicative parameter, modifying    
  the phenomenological loop production term of equation    
  (\ref{gdtsimple}). A numerical study of the functional   
  dependence of $P_{\rm eff}$ on $P$ in an expanding universe   
  will be presented in \cite{inpreparation}.

  The second possibility arises from the observation that the 
  velocities in the three infinite dimensions and in the 
  compact ones, $v_x$ and $v_\ell$ respectively, satisfy the   
  constraint ${v_x}^2+{v_\ell}^2\le 1/2$. Together with the fact   
  that $v_\ell$ is very weakly damped compared to $v_x$ (section  
  \ref{solutions}), this will slow down the motion of the strings    
  in the three infinite dimensions. It is even possible that string  
  motion in these dimensions can completely stop, in which case the   
  strings will no longer be able to intercommute. We study this effect   
  in more detail below.

 \subsection{\label{vosbrane}Applying the EDVOS Model}         

  \begin{figure}
   \includegraphics[width=2.5in,keepaspectratio]{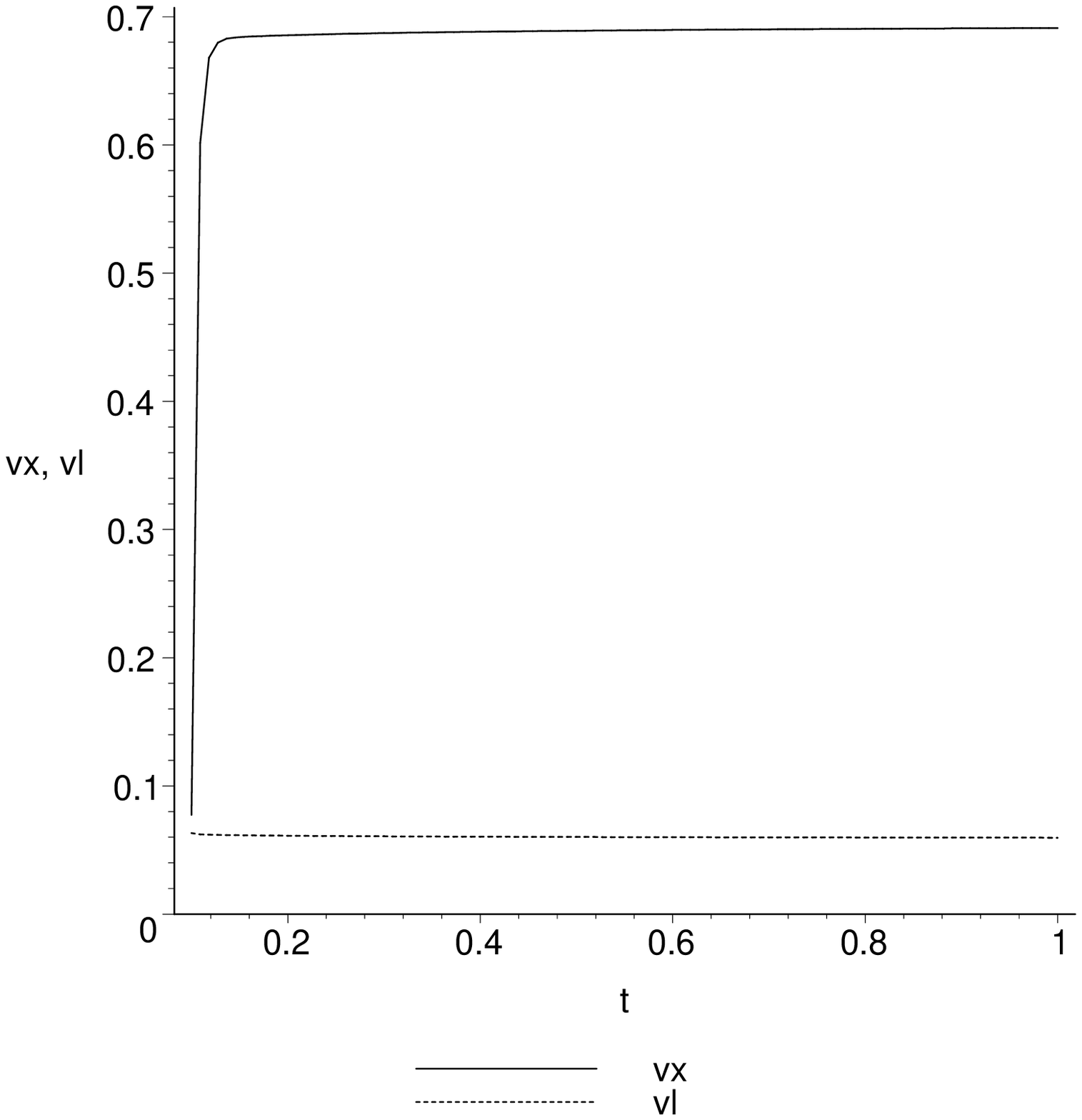}
   \includegraphics[width=2.5in,keepaspectratio]{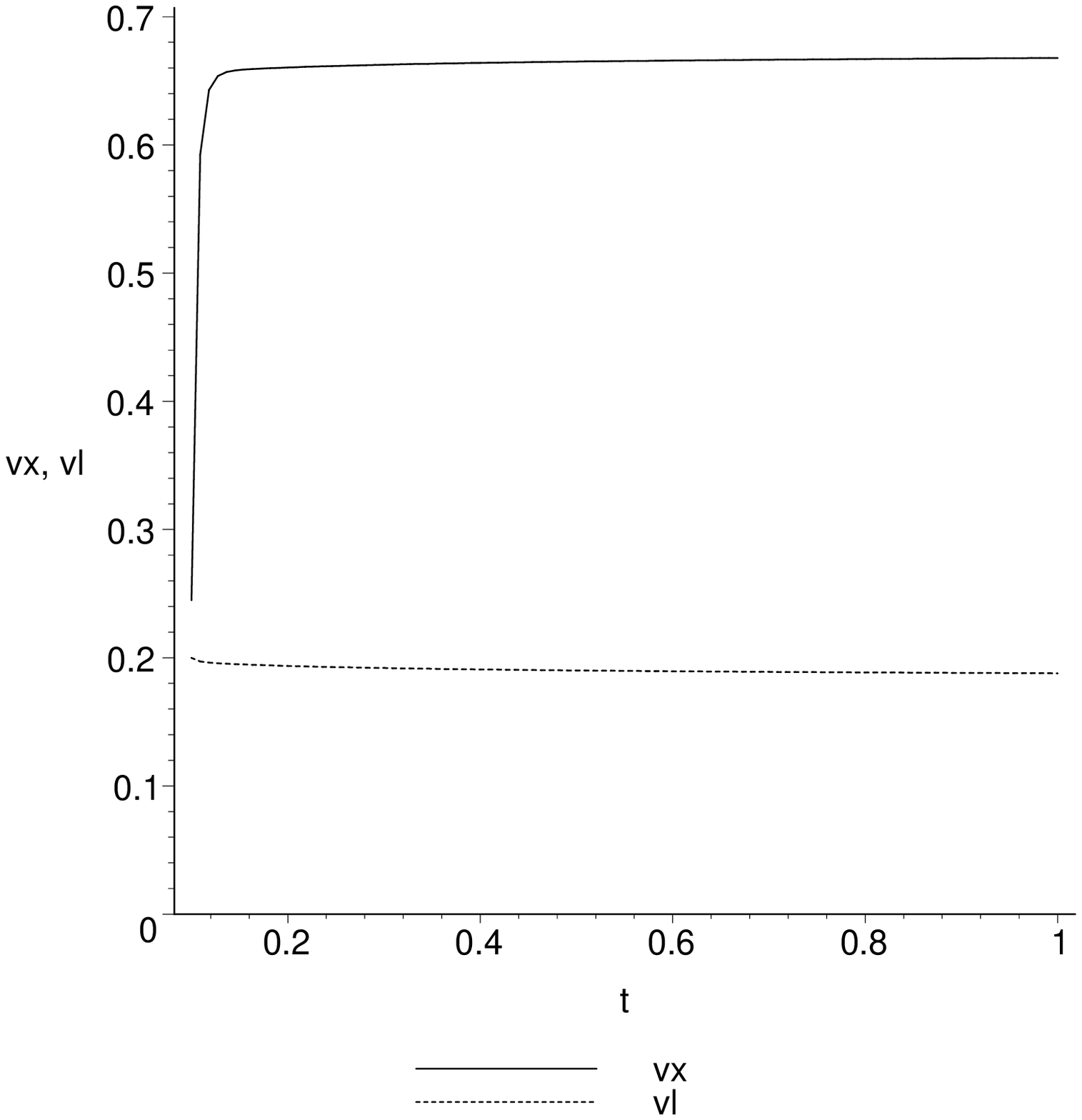}
   \includegraphics[width=2.5in,keepaspectratio]{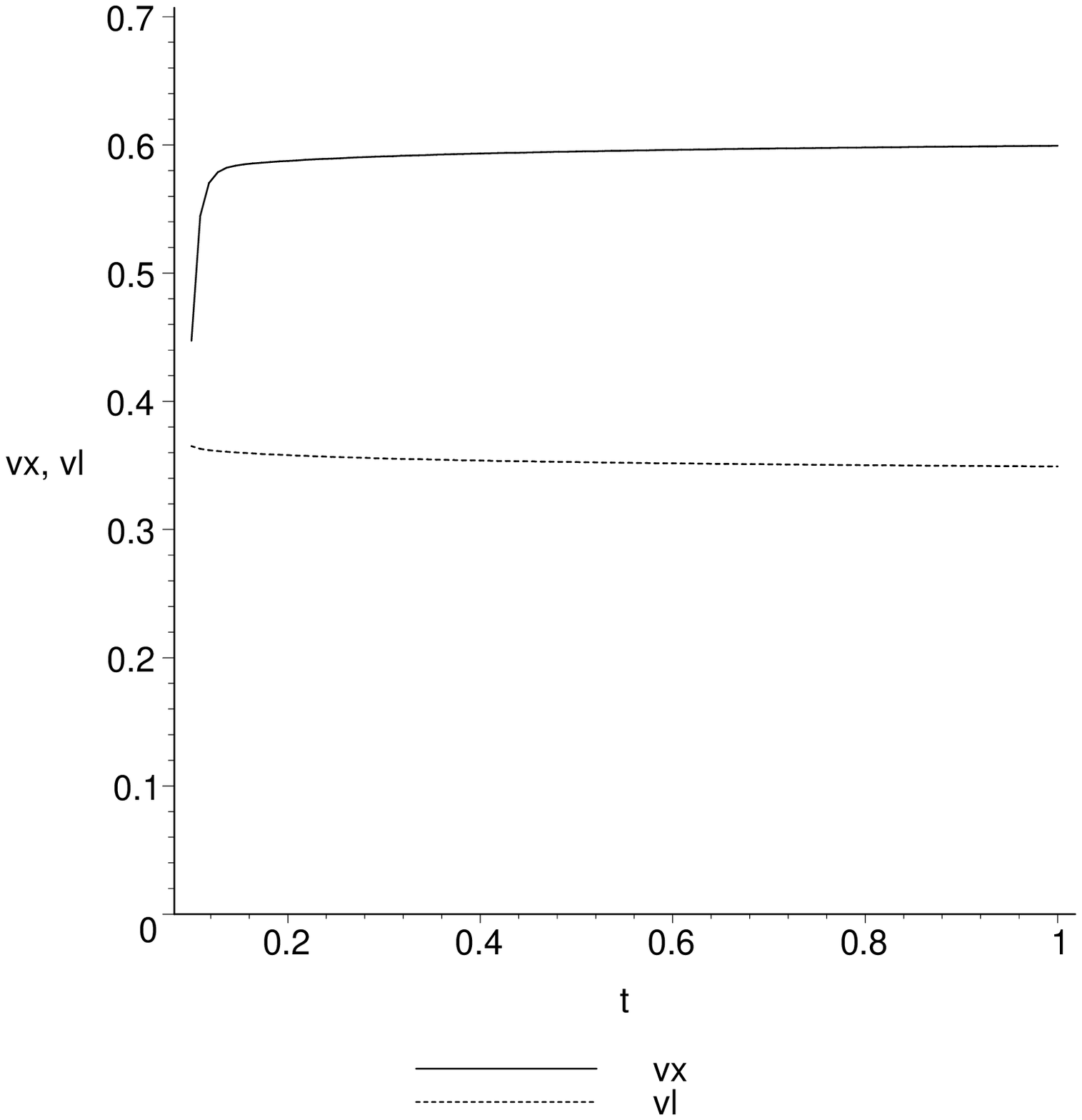}
   \includegraphics[width=2.5in,keepaspectratio]{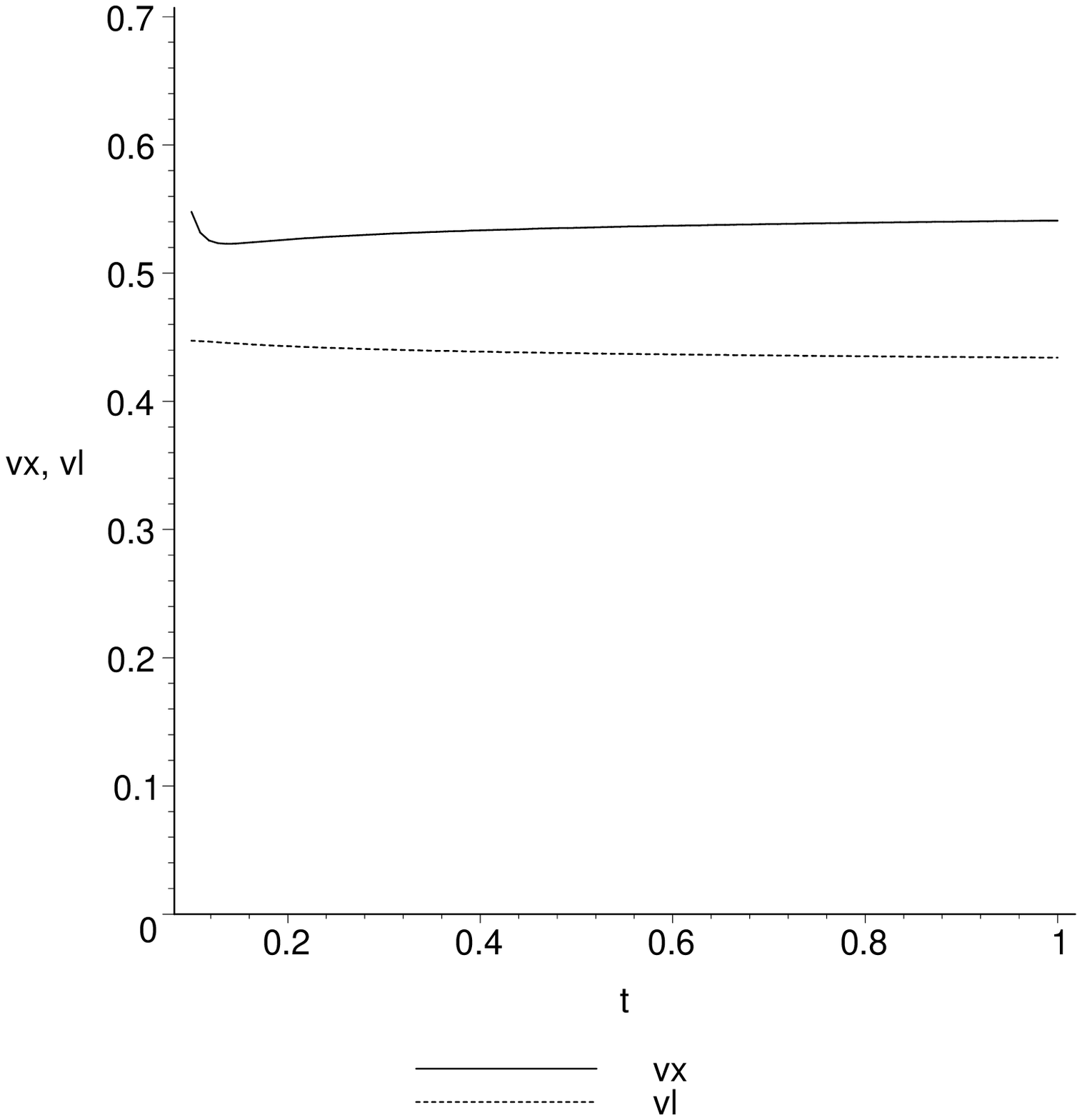}
   \caption{\label{v0var} Velocity evolution in arbitrary time units
            with initial condition for the total velocity $v_0=0.1,1/
            \sqrt{10},1/\sqrt{3}$ and $1/\sqrt{2}$. We have
            assumed equipartition of kinetic energies at $t=0$ and
            that the number of dimensions transverse to the brane is 2.
            For small $v_{\ell 0}$ the three dimensional velocities scale
            to a value close to the purely three-dimensional result $v\simeq
            0.7$. However for $v_{\ell 0}>0.4$ the value of $v_x$ can be
            significantly less than $0.6$.}
  \end{figure}

  \begin{figure}
   \includegraphics[width=3in,keepaspectratio]{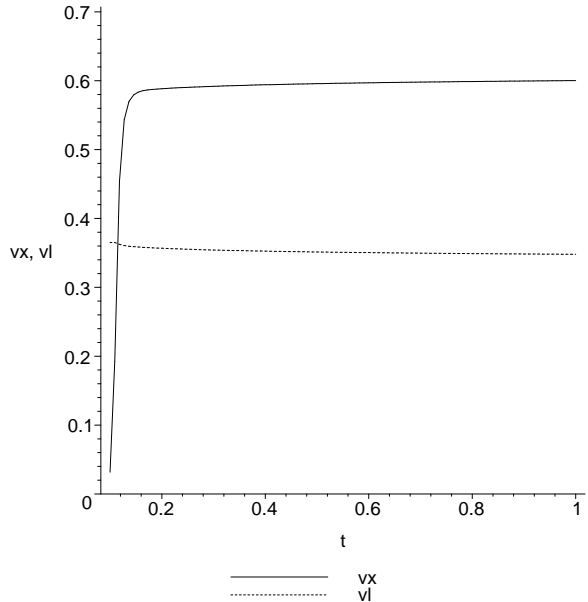}
   \caption{\label{vx0} Velocity evolution in arbitrary time units
            for $v_x \ll v_\ell$ at $t=0$. Here, $v_x$ rapidly increases and
            reaches the same equilibrium value as in the case $v_x \sim
            v_\ell$ at $t=0$. This shows that the assumption of
            equipartition of kinetic energies at string formation is
            not necessary. The important initial condition is $v_\ell$.}
  \end{figure}

  \begin{figure}
   \includegraphics[width=3in,keepaspectratio]{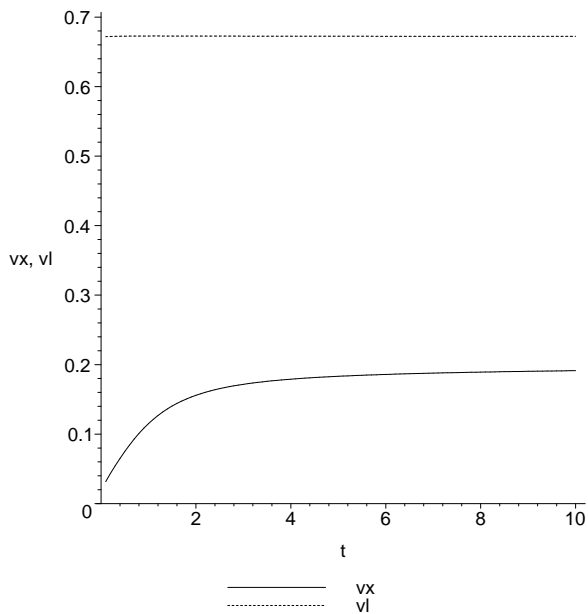}
   \caption{\label{vllarge} Velocity evolution in arbitrary time units
            for $v_\ell=0.95/\sqrt{2}$ at $t=0$. Because $v_\ell$
            is so high, $v_x$ scales to a low value of about $0.2$.
            Such a low velocity also affects the string density
            (Fig. \ref{gammavl}).}
  \end{figure}

  \begin{figure}
   \includegraphics[height=2.7in,width=3in]{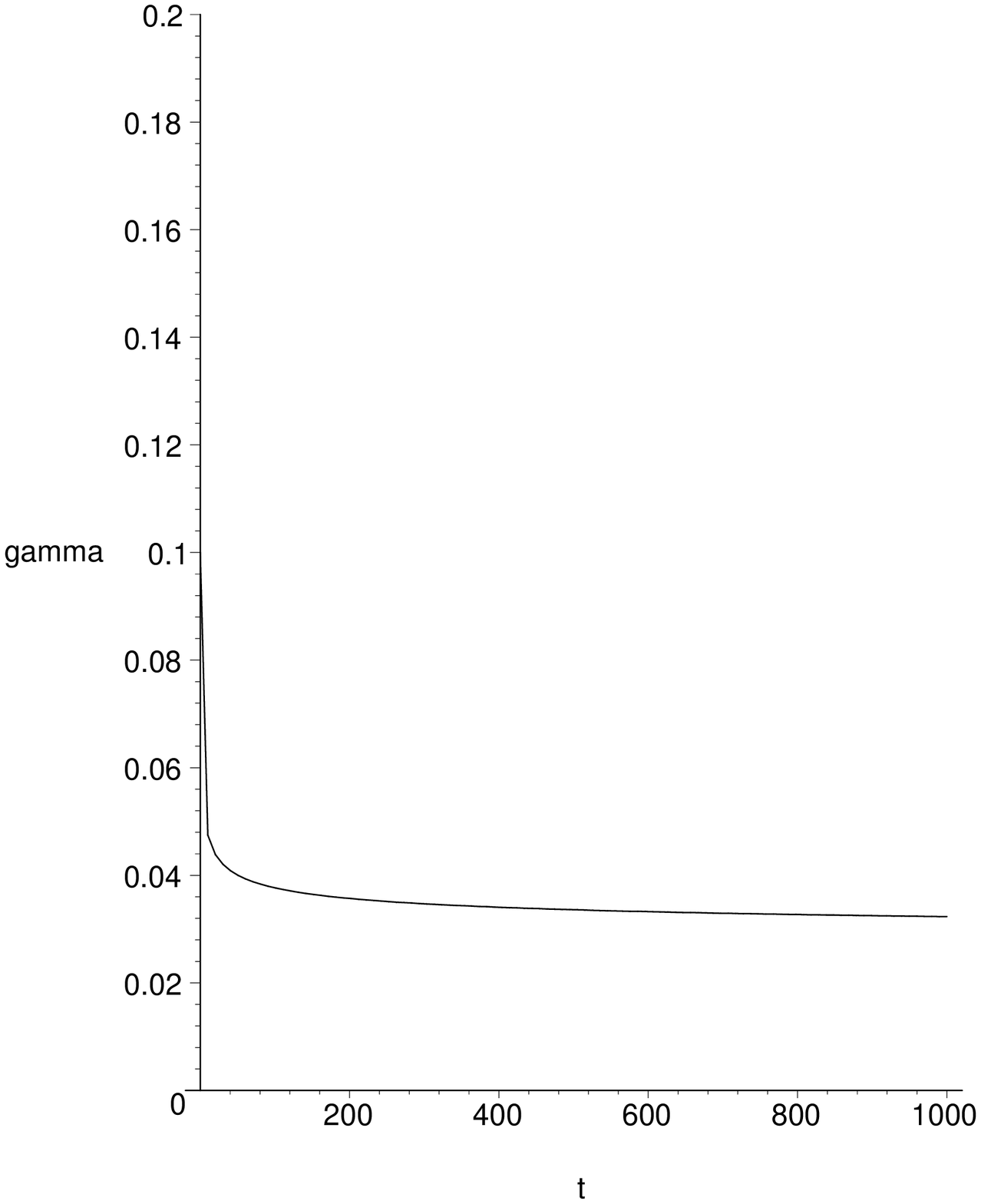}
   \includegraphics[height=2.7in,width=3in]{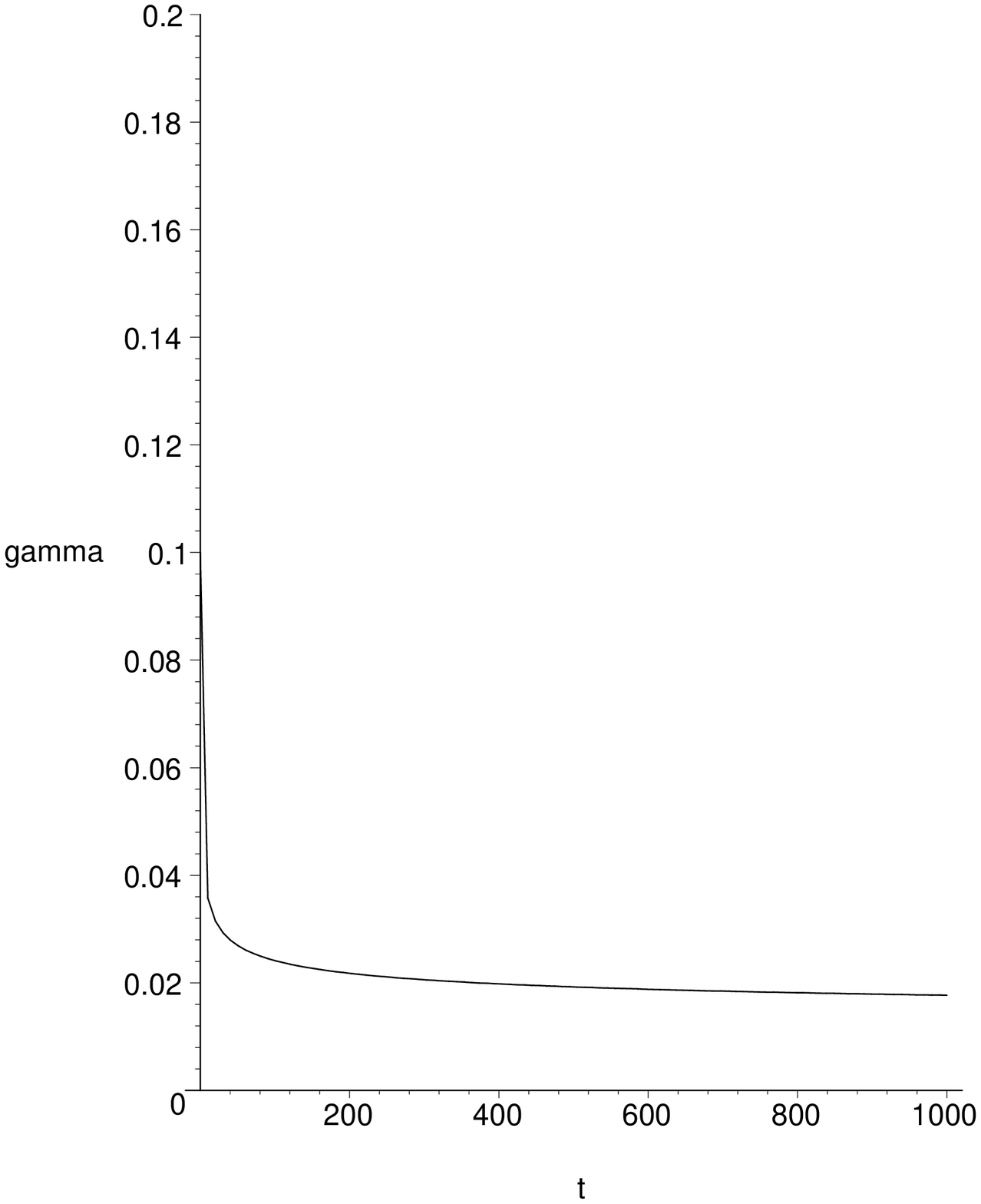}
   \caption{\label{gammavl}The effect of slowing down of strings
            on $\gamma$. On the left is the evolution of $\gamma$
            for $v_\ell \simeq 0.3$ at formation. On the right is
            the corresponding evolution for $v_\ell=0.95/\sqrt{2}$,
            in which $v_x$ is considerably smaller.}
  \end{figure}
 
  As we have seen, the spatial structure of the string network 
  at formation is essentially three-dimensional, but the strings   
  evolve in $9-p+4$ spacetime dimensions with $9-p$ of them 
  compactified. The compactification radius is assumed to  
  be stabilised at a size smaller than the horizon \cite{DvalTye}.    
  Thus we can model the evolution of the string density 
  using an effective three-dimensional description, 
  where we introduce an intercommuting probability $P$ 
  to account for the fact that strings can `miss' when   
  they evolve in higher than $4$ spacetime dimensions.  
  As an evolution equation for $\gamma$ we therefore use   
  (\ref{gdt}) and replace the loop production term $\tilde c   
  v_x {\gamma}^{-1}$ by $P_{\rm eff} \tilde c v_x {\gamma}^{-1}$,  
  where an {\it effective} intercommuting probability $P_{\rm eff}=f(P)$  
  has been used to implicitly take into account the effect of string   
  wiggliness on the scaling dependence on $P$. We also need  
  to include  both $v_x$ and $v_\ell$, the velocities in the   
  $(3+1)$-dimensional FRW $x$-space and the $(9-p)$-dimensional    
  compact $\ell$-space respectively, which we evolve using   
  equations (\ref{v_xdt}),(\ref{v_ldt}). The parameters $k_x$   
  and $k_\ell$ are given by (\ref{k_xans}-\ref{k_lans}) for   
  $D=9-p+3$ with $\frac{v_{xc}}{v_c}\simeq \frac{v_x}{v}$ and   
  $\frac{v_{\ell c}}{v_c}\ll 1$ (see section \ref{solutions}).

  The formation of the network takes place on the plane of the colliding
  branes, which has a finite thickness. One may worry that this fuzziness  
  will translate to an uncertainty in the position of the strings in the   
  extra dimensions, which may give rise to a significant structure in   
  the $\ell$-space just after the strings are formed. This could spoil the  
  assumption that the initial structure of the network can be considered  
  three-dimensional, in which case equation (\ref{gdt}) would not be valid.     
  Fortunately this thickness is of the order of the inverse superstring    
  scale $M_s^{-1}$, which is approximately $10^4$ times smaller than  
  the correlation length at formation (see below). The network is to  
  a very good approximation three-dimensional even at the time of formation.  
  This justifies the assumption $k_\ell \ll k_x$ or equivalently   
  $\frac{v_{\ell c}}{v_c}\ll 1$ at early times.  

  The correlation length at formation, estimated by studying the   
  tachyon potential, is of the same order of magnitude as the 
  expected horizon size at that time \cite{SarTye} 
  \be\label{L0}   
   L_0 \sim H^{-1} \sim 10^4 \; M_s^{-1} \; ,    
  \ee    
  where the superstring scale is set by CMB data to a low GUT scale  
  \be\label{Ms}   
   M_s \sim 10^{14} \; \rm{GeV}.  
  \ee   
 
  After the strings are formed, the correlation length grows because    
  of the Hubble expansion. On the other hand the dimensions transverse  
  to the brane are compactified at a size a few times the inverse    
  superstring scale, which is some $10^3$ times smaller than the correlation  
  length. Thus, if the initial long string network is Brownian, it will  
  remain so (as in the usual three-dimensional case) and equation (\ref{gdt})   
  will be valid at all times.           
 
  To simplify the equations we can use the ansatz $w_\ell^2=  
  \frac{v_\ell^2}{v^2}$ (section \ref{simplemodel}). Equations    
  (\ref{gdt}),  (\ref{v_xdt}) and (\ref{v_ldt}) are then replaced by   
  \bq      
    && \gamma^{-1}\,\frac{d\gamma}{dt}=\frac{1}{2\,t} \left[\beta \left(
        2+2{v_x}^2+\frac{{v_\ell}^2}{v^2} \right) -2 + \frac{P_{\rm eff}   
        \tilde c v_x}{\gamma}\right]   
       \label{gdtbrane}\\
    && v_x \frac{d v_x}{dt}=\frac{1}{t}\left[\frac{k_x v_x}{\gamma}
        (1-v^2) - \left(2-2{v_x}^2-\frac{{v_\ell}^2}{v^2}\right)
        \beta{v_x}^2 \right]
       \label{v_xdtbrane}\\
    && v_\ell \frac{d v_\ell}{dt}=\frac{1}{t}\left[\frac{k_\ell v_\ell}
        {\gamma}(1-v^2)-(1-2v^2)\left(1-\frac{{v_\ell}^2}{v^2}\right)
        \beta{v_\ell}^2 \right]
       \label{v_ldtbrane}
  \eq   
  where we have explicitly written the loop production parameter in terms  
  of the effective intercommuting probability $P_{\rm eff}$, that is, we   
  have set   
  \be
   \tilde c \rightarrow P_{\rm eff} \tilde c \; .     
  \ee  
 
  Below we discuss numerical solutions of equations 
  (\ref{gdtbrane}-\ref{v_ldtbrane}) with different initial   
  conditions for $v_x$, $v_\ell$ as well as different values for the    
  effective intercommuting probability $P_{\rm eff}$ in the range   
  between $10^{-3}$ and $1$. For an illustration we first consider   
  the case $P_{\rm eff} \simeq 0.1$. As in section \ref{solutions}   
  we have that since $\frac{v_{\ell c}} {v_c}$ is small, there is   
  essentially no source term for $v_\ell$ in (\ref{v_ldtbrane})   
  and thus $v_\ell$ is just given by its initial condition slowly    
  damped by the expansion. On the other hand $v_x$ is sourced by    
  the string curvature $R\sim L$ and although it is more strongly    
  damped it dominates $v_\ell$. There is a critical value of   
  $\frac{v_{\ell c}}{v_c}\simeq 0.15$ above which the curvature (source) 
  term for $v_\ell$ becomes large enough for $v_\ell$ to dominate.  
  This value is too large to be relevant in the context of brane 
  inflation: it corresponds to a situation where more than one tenth  
  of the velocity developed during the last correlation time is in 
  the $\ell$-dimensions, which would require the curvature vector to 
  have a significant component in the $\ell$-space. As explained above    
  the correlation length (the typical radius of curvature of strings)   
  at formation is of order $10^3$ times the size of the extra 
  dimensions and further grows with the expansion, so    
  $\frac{v_{\ell c}}{v_c}$ is expected to be much less than this  
  critical value. Thus, string propagation in the $x$-space will not  
  stop, but it can be significantly slowed down if the strings are  
  created with enough momentum in the extra dimensions.   

  Indeed we expect that some of the energy associated with the brane  
  collision will be converted to kinetic energy of string segments in  
  the dimensions transverse to the brane, so the slowing down of string  
  motion in the infinite dimensions could be a significant effect. In Fig.   
  \ref{v0var} we plot $v_x$ and $v_\ell$ as functions of time, assuming  
  equipartition of kinetic energies at formation. This assumption is not  
  necessary, and even if $v_x \ll v_\ell$ initially, $v_x$ will increase    
  very fast (Fig. \ref{vx0}) because of the curvature source term in   
  equation (\ref{v_xdtbrane}). As long as the constraint $v^2 \le 1/2$ is   
  satisfied, the important initial condition is $v_\ell$ at the time of  
  string formation. From Fig. \ref{v0var} we see that for initial values 
  of $v_\ell$ smaller than 0.4 the scaling value of $v_x$ is less than a  
  few percent different from the purely three-dimensional result $v\simeq 0.7$.  
  However for $v_\ell$ initially greater than this value, the slowing   
  down of strings in the three infinite dimensions is significant.   
  In particular if $v_\ell$ has a value close to the maximum allowed   
  ($v_\ell \simeq 1/\sqrt{2}$) then $v_x$ will approach a very small   
  value $v_x \simeq \sqrt{1/2-{v_\ell}^2}$ (Fig. \ref{vllarge}). Such a   
  dramatic reduction of the three-dimensional speed of the strings   
  also has a significant effect on the scaling value of $\gamma$ (Fig.  
  \ref{gammavl}) because the loop production term is proportional to   
  $v_x$: slowly moving strings will intercommute less often so the  
  final density of the network will be higher, corresponding to a   
  smaller $\gamma$.                                 
 
  It is interesting to compare our results with the discussion of   
  Jones, Stoica and Tye \cite{JoStoTye2}. As a first approximation   
  they used a simple one-scale model and assumed a relativistic,    
  constant speed of strings. They found that the scaling value of 
  the energy density of the network is $P^{-2}$ times greater than   
  in the purely three-dimensional case. The leading correction to this   
  result comes from allowing the velocity of strings to be a    
  variable, that is using a VOS model rather than a simple one-scale    
  model. This is equivalent to setting $v_\ell=0$ in our extra   
  dimensional EDVOS model. The effect of this variable velocity   
  is to reduce the density by a factor of 10 or so. We then include    
  velocities in the extra dimensions $v_\ell \neq 0$. If $v_\ell<0.4$   
  there is no additional observable effect.  If $v_\ell > 0.4$ but    
  considerably smaller than $1/\sqrt{2}$ then these extra dimensional     
  velocities have no significant effect on the energy density, though    
  they may lead to a substantial reduction of the three-dimensional    
  string velocities. Finally, if $v_\ell \simeq 1/\sqrt{2}$ the    
  strings will be moving very slowly in the three infinite dimensions  
  so that the number of intercommutings will be further reduced. This    
  results in a substantial increase of the energy density of the    
  string network.      

\begin{figure}
   \includegraphics[height=1.8in,width=2in]{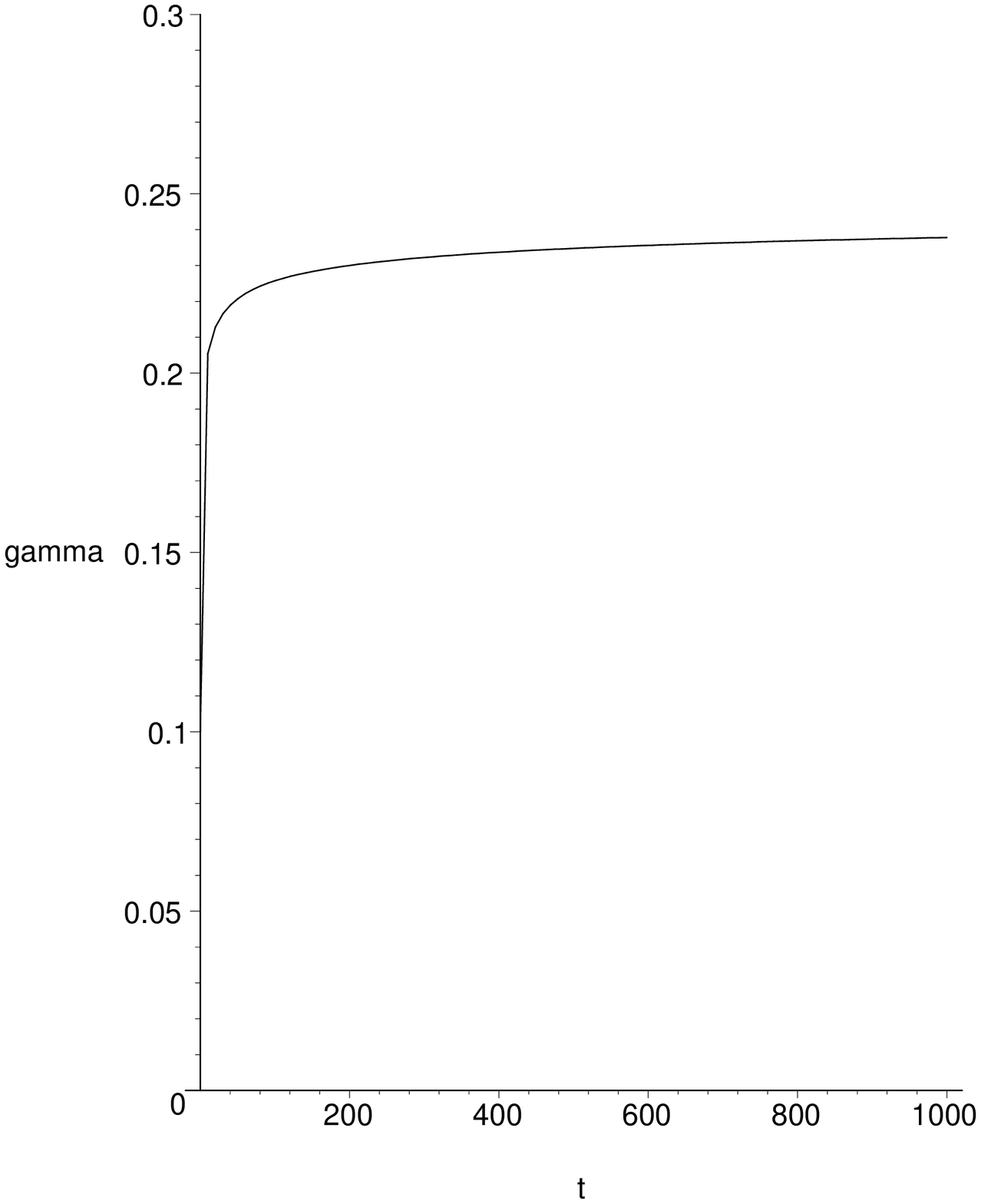}
   \includegraphics[height=1.8in,width=2in]{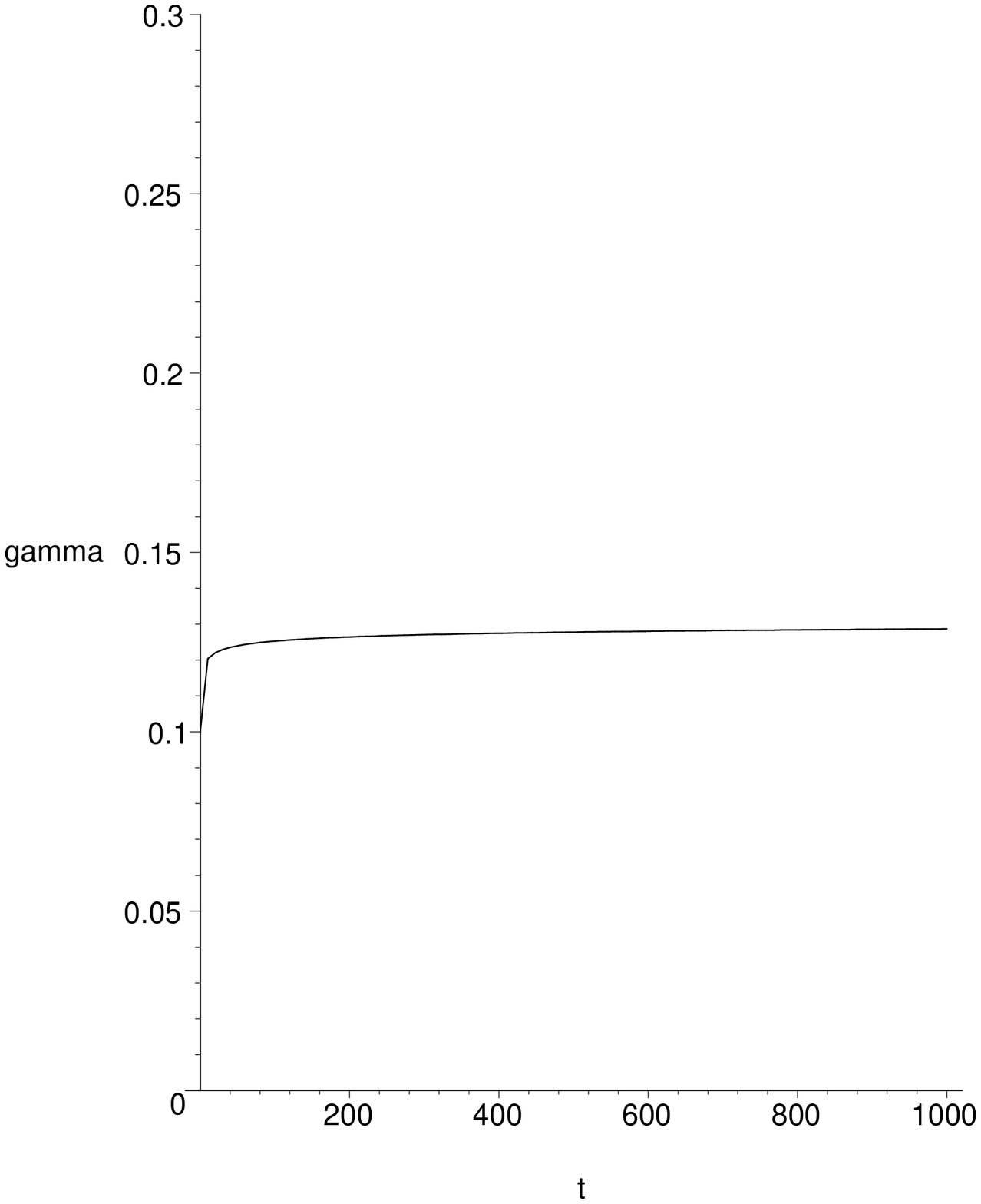}
   \includegraphics[height=1.8in,width=2in]{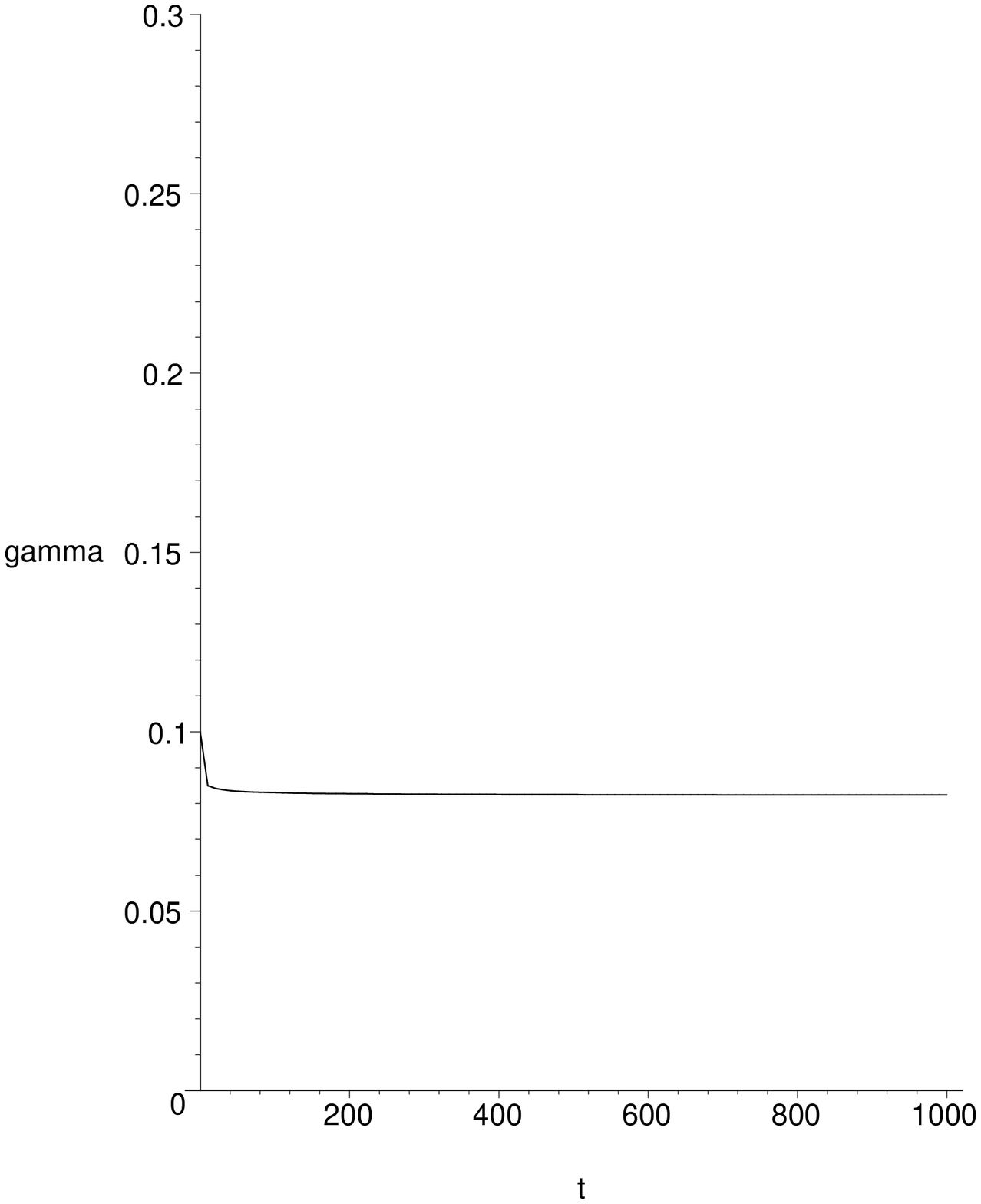}
   \includegraphics[height=1.8in,width=2in]{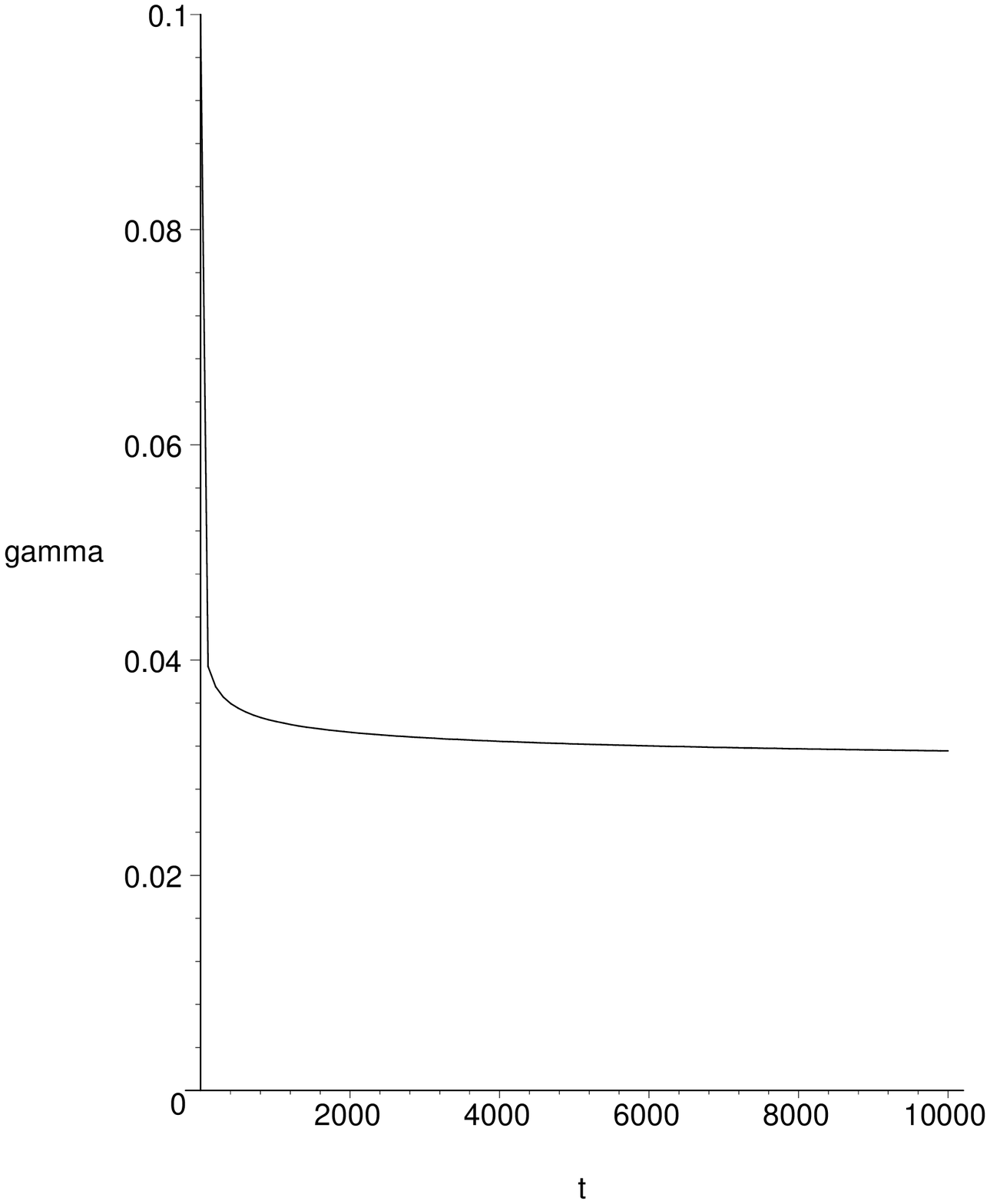}
   \includegraphics[height=1.8in,width=2in]{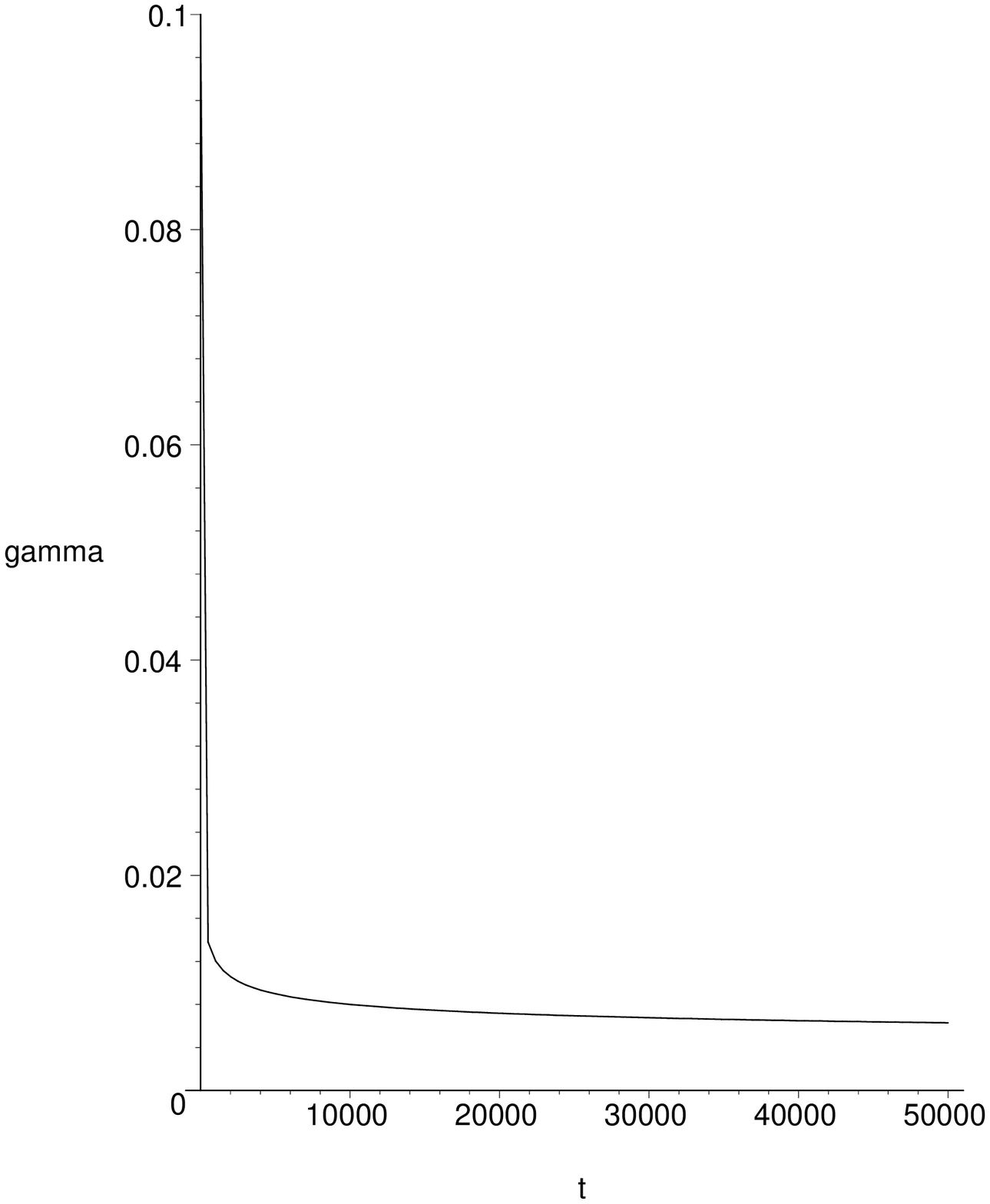}   
   \includegraphics[height=1.8in,width=2in]{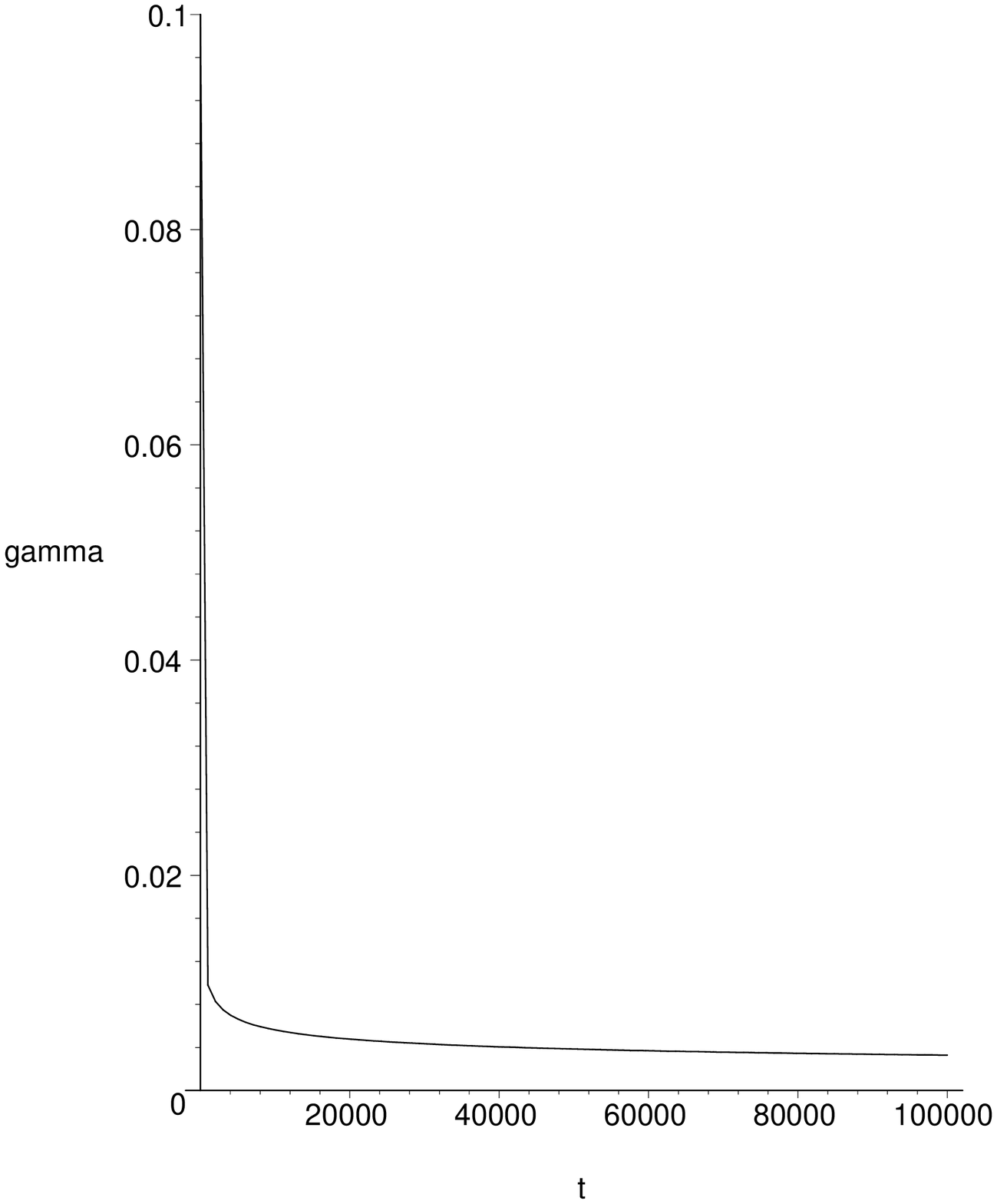} 
   \caption{\label{gofP} The dependence of $\gamma$ on the 
            effective intercommuting probability $P_{\rm eff}$.   
            From top left to bottom right $P_{\rm eff}=1,0.5,0.3,  
            0.1,0.01,10^{-3}$. As $P_{\rm eff}$ tends to zero the   
            approximation $\gamma\simeq P_{\rm eff}$ becomes better   
            but the time required to reach scaling increases.}   
  \end{figure}

  Finally we study the effect of varying the intercommuting 
  probability $P$. For networks of the same type (F or D-string 
  networks), $P$ is expected to be in the range $10^{-3}<P<1$ 
  (in particular $10^{-3}<P<1$ for F-strings and $0.1<P<1$ for  
  D-strings \cite{PolchProb}). However, since the strings can  
  develop significant small-scale structure and become wiggly,  
  they may have more than one opportunity for reconnection in  
  each crossing time. Our one-scale model does not fully
  take into account such wiggly effects, but we have done so  
  implicitly by using the effective intercommuting probability   
  $P_{\rm eff}=f(P)>P$, as we have discussed above. The effect   
  of small-scale wiggles can only be accounted for numerically   
  and the results of such studies in an expanding space will be   
  presented in \cite{inpreparation}. For the present discussion  
  we take $P_{\rm eff}$ in the range $10^{-3}\le P\le 1$ and    
  study how the model is affected by changes in $P_{\rm eff}$.    
  In Fig. \ref{gofP} we show the behaviour of $\gamma$ for    
  different values of $P_{\rm eff}$ in that range assuming a    
  moderate $v_\ell\simeq 0.36$ at formation. Reducing $P_{\rm eff}$  
  leads to less intercommutings and therefore greater string energy   
  density, corresponding to a smaller $\gamma$. 
   
\section{\label{conc}Summary and Discussion}   
  
 We have presented an extension of the VOS model that can be used to study   
 the macroscopic evolution of a Brownian network of Nambu-Goto cosmic   
 strings in cosmological spacetimes with extra dimensions (EDVOS model).   
 In order for the Brownian structure of the network to be preserved by   
 time evolution either the spacetime must be isotropic or, in the case of  
 anisotropic expansion, the network must be formed on an isotropic slice   
 of the spacetime. If the strings are not confined on this slice after   
 their formation then the extra dimensions must be compactified at a size   
 smaller than the correlation length, as is the case in models of brane 
 inflation. The evolution has then an effective, three-dimensional   
 description, in which the effect of possible velocities in the extra   
 dimensions can be taken into account. This is a significant factor   
 because extra-dimensional velocities will act to slow down string   
 motion in the infinite dimensions, reducing the number of string   
 intercommutings while also changing the strings' effective 3D 
 energy per unit length.  

 First, we applied the model to the case of an isotropic ($D+1$)-dimensional   
 FRW universe and found the generic behaviour (for $D>3$) in which   
 $L\propto a^{D/(D-1)}$ (in contrast to the naive conformal stretching   
 obtained without the EDVOS model $L \propto a$). Obviously, these strings   
 do not scale because they find it increasingly difficult to find each   
 other and intercommute, and so they would quickly dominate the energy   
 density of such a universe. However, even if the additional dimensions   
 were compact, this evolution might pertain at early times, as long   
 as the expansion remained isotropic. In this case, the network    
 correlation length $L$ would continue to expand until it inevitably   
 catches up with $R_\ell \propto a$, the scale of the expanding compact   
 dimension. Thereafter, the network would become effectively    
 three-dimensional and, if the compact dimensions were then stabilised,  
 the evolution would turn over towards the alternative FRW 3-brane 
 scenario we discussed earlier. We point out therefore, in the case of 
 static compact extra dimensions that, even if we begin with an initial 
 correlation length smaller than the compact dimension $L_{\rm 0}< R_\ell$,   
 the correlation length will grow until $L>R_\ell$, at which point the    
 evolution becomes effectively three-dimensional. Hence, we conclude   
 that the usual assumption $L_0\gg R_\ell$ is not actually necessary for   
 achieving asymptotically 3D scaling evolution.

 The most pertinent application of our model is to a situation where the   
 string network is formed on a FRW 3-brane with the dimensions transverse  
 to it compact and small (we have only considered flat extra dimensions).   
 We have allowed the strings to be able to explore the bulk after formation,   
 as is the case in models of brane inflation. We find that the density of   
 the network in our VOS model (after scaling has been achieved) can be up    
 to a factor of 10 smaller than previous estimates \cite{JoStoTye2}. This   
 correction comes from quantitatively accounting for the role of the    
 string 3D velocity, allowing it to be a variable rather than a constant   
 as in the simplest one-scale models. We also found additional effects    
 arising from the fact that the strings can also move in the extra    
 dimensions. Standard field theory strings can be shown numerically to   
 approach an average velocity of about $1/\sqrt{2}$ (on lengthscales   
 $\lambda \ll H^{-1}$). This is also the case in our extra-dimensional   
 VOS model but now a significant amount of this velocity can be trapped   
 in the extra dimensions, with the result that the observable 3D velocity  
 is reduced. Since only the three FRW dimensions are expanding while the   
 size of the extra dimensions is stabilised, there will only be a very  
 weak redshifting of the extra-dimensional velocities. Thus, if the   
 strings are created with significant velocities in the extra dimensions,    
 these could survive for a long time and result in a slowing down of  
 string motion in the three large dimensions.    

 One could argue that after reheating, the strings would enter a damping  
 regime where velocities would slow down due to the damping force from  
 a high radiation background density. This could damp velocities in the   
 extra dimensions more strongly than Hubble expansion, in particular
 eliminating $v_\ell$. However, we note that reheating only takes place on  
 the brane (the bulk remains cold and empty \cite{DvalTye}) and it is only   
 the small portion of the string network intersecting the brane which feels   
 this damping. As the strings move, they briefly encounter the brane and  
 pass through it, feeling a damping force, but at any time most of the    
 length of the network is in the bulk, where no friction is felt. There  
 are only a finite number of such brief encounters before cosmological  
 expansion on the brane cools it sufficiently for the friction force to   
 become negligible. The strings also might be oriented such as to intersect   
 and remain in contact with the 3-brane in the longer term, but this only   
 implies that the string will be `pinned' at one point. Taking into account    
 these geometrical effects, it appears that the usual 3D frictional damping    
 terms will be suppressed by factors of order the intercommuting probability 
 $P$ (for both $v_\ell$ and $v_x$).  The network in higher dimensions
 will approach its relativistic scaling regime much more quickly than its
 3+1 FRW counterpart, though a more quantitative investigation is   
 certainly warranted.   

 Gravitational back-reaction can also be expected to reduce the   
 extra-dimensional velocities. The effect of gravitational backreaction    
 can be incorporated in the VOS model by including a term $8\Gamma G \mu   
 v^6$ in the evolution equation for the correlation length $L$ \cite{vosk},   
 where for long strings $\Gamma$ is a constant of order $10$ \cite{vosk}.    
 Even if $G \mu$ has a value close to the upper limit $10^{-6}$, this term    
 is small compared to the loop production term $P \tilde c v$ so it has    
 little direct influence on the network density. However, its effect on   
 the extra-dimensional velocity $v_\ell$ could be more dramatic since    
 gravitational radiation tends to act to eliminate small-scale structure   
 below a certain length-scale $\lambda$.  This lengthscale was previously   
 thought to be $\lambda \sim \Gamma G\mu t$, but closer investigation has    
 suggested that it was in fact considerably smaller $\lambda \sim    
 (\Gamma G\mu)^{\eta} t$ with $\eta = 3/2$ in the radiation era and $\eta    
 = 5/2$ in the matter era \cite{SieOlVil}. The reanalysis demonstrated    
 that only modes of comparable wavelength interact efficiently, a fact   
 which may be relevant in the case with extra dimensions. Here, we could    
 envisage the small-scale modes trapped in the extra dimensions essentially 
 decoupling from the effective 3D evolution.  Damping of the velocity   
 $v_\ell$ then would be primarily through self-interactions between the  
 trapped modes which tend toward slow power law, rather than exponential,    
 suppression.  Whatever the outcome of a closer examination of this issue   
 in higher dimensions, it is clear that the extra dimensional velocity    
 $v_\ell$ will have a long lifetime and will influence the network evolution  
 over long timescales. Note however that since the string position in the   
 compact dimensions is described by (worldsheet) scalar fields, one expects  
 some stabilisation mechanism for these moduli to kick in at low energies.   
 This would render these excitations massive, effectively localising the   
 string in the extra dimensions \cite{PolchStab}. Given the large  
 hierarchy between the preferred inflationary and SUSY breaking scales   
 in brane inflation models (GUT and TeV respectively), the extra-dimensional    
 effects we have considered are likely to be relevant for many orders of    
 magnitude in time.       

 While our EDVOS model can characterise the effect of the velocity $v_\ell$   
 on the overall large-scale 3D network properties, uncertainties remain   
 as to its magnitude at network formation, the subsequent (weak) rate   
 of damping, and whether it can be sourced in any way during the    
 subsequent evolution. Given the brane collision out of which the string   
 network forms, it seems plausible that the string network will be    
 released from the brane with a significant velocity $v_\ell$ with which   
 to traverse the extra bulk dimensions.  If $v_\ell$ is relativistic,   
 then the average 3D velocity $v_x$ will be smaller and the network   
 density will be higher than previously expected. However, this presumes  
 the absence of any significant source terms for $v_\ell$, for example,   
 due to string reconnections or other dynamical effects. By including   
 such source terms, we have demonstrated that there is, in principle,   
 the pathological possibility that the Hubble-damped 3D velocity $v_x$   
 becomes negligible because of continuous contributions to the undamped  
 $v_\ell$.  In this case the string network would dominate over radiation  
 in the early universe.  But we have also shown that there is a threshold  
 ($k_\ell \sim 0.1k_x$) below which this does not happen. For brane   
 inflation scenarios in which the string curvature is predominantly    
 three-dimensional at formation, we expect this criteria to be    
 fulfilled, with the EDVOS model predicting a scaling solution in which  
 $L = \gamma t$ and $\gamma\sim P_{\rm eff}$. We stress that the statement  
 $L \propto P$ should not be taken as a prediction of our model. Instead, we  
 have used an {\it effective} intercommuting probability $P_{\rm eff}$, which   
 is a phenomenological parameter of the model, to be determined by string 
 network simulations or a deeper statistical analysis. The functional   
 dependence of $P_{\rm eff}=f(P)$ on the {\it actual} intercommuting   
 probability $P$ will be the subject of a forthcoming publication   
 \cite{inpreparation}.       

 Another interesting issue which we can use our EDVOS model to investigate   
 is the evolution of closed strings (loops), and in particular the   
 possibility that such loops could wrap around the compact dimensions   
 \cite{Vilenk}. If the compact manifold admits non-trivial one-cycles,   
 this would give rise to stable monopole-like objects (from the 3D point   
 of view) which could dominate the universe or even provide a dark matter   
 candidate. We denote these objects {\it cycloops}, that is, loops wrapping   
 around non-trivial cycles. We leave the study of the formation, properties  
 and consequences of cycloops for a different publication \cite{inprep1}. 
 
 The purpose of the present paper was not to exhaustively explore
 all possible avenues or to investigate specific brane inflation 
 models, but rather to provide a broad picture of the key dynamical 
 effects which will influence a cosmic string network emerging from a   
 higher dimensional theory. We believe the generalised VOS model we    
 have presented here is an important step in developing a more     
 quantitative description of cosmic string evolution in brane    
 inflation and other contexts. There are a host of further issues    
 to explore. We have, for example, only considered the `abelian'    
 case where there is only one particular type of string, which   
 simply exchanges partners and reconnects when it meets another (albeit  
 with a reduced probability for such an encounter). Similar methods can  
 be applied to the richer structures created by $F$ and $D$ strings in   
 more complex (true) string networks (see Ref. \cite{MartNonInt}) and    
 this deserves further investigation.

\appendix  
 \section{\label{appendix}Approximate Formulae for $k$, $k_x$ and $k_\ell$} 

  Defining $\dot{\bf y}\!=\!(\dot{\bf x},\dot{\bf l}/a)$ we split the  
  velocity $\dot{\bf y}$ into a `curvature' component $\dot{\bf y}_c$  
  produced during the last correlation time and a `left-over' component  
  $\dot{\bf y}_p$ (coming from previous accelerations) by writing 
  $\dot{\bf y}\!=\dot{\bf y}_c + \dot{\bf y}_p$. We interpret the 
  first as the velocity induced on large-scales by the present 
  correlation length of the string, whereas the second is the velocity 
  remaining from previous correlation times, generally on small scales.
  These two components  
  are uncorrelated so we have $\langle \dot{\bf y}_c \cdot \dot{\bf y}_p    
  \rangle = 0$. We also have the gauge condition $\dot{\bf y}    
  \cdot{\bf y}^\prime\!=\!0$ and we assume that both components   
  $\dot{\bf y}_c$ and $\dot{\bf y}_p$ separately satisfy this, i.e.   
  we have $\dot{\bf y}_c\cdot{\bf y}^\prime=0$ and $\dot{\bf y}_p
  \cdot{\bf y}^\prime=0$.

  Hence, since the curvature vector $\bf u$ is normal to the tangent
  $\bf{y}^\prime$ we have
  \be\label{uhat}
   \hat{\bf u} = A\hat{\dot {\bf y}}_c - B\hat{\dot {\bf y}}_p + 
   \sum_{i=1}^{D-2} C_{i} \hat{\bf y}_{i}
  \ee
  where $\hat{\dot {\bf y}}_c\:$, $\hat{\dot {\bf y}}_p$ are the unit  
  vectors in the direction of $\dot{\bf y}_c\:$, $\dot{\bf y}_p$ 
  respectively and $\hat {\bf y}_{i}$ are unit vectors spanning the  
  $(D-3)$-dimensional subspace normal to $\dot{\bf y}_c\:$,   
  $\dot{\bf y}_p$ and ${\bf y}^\prime$. Dotting with $\dot{\bf y}$  
  and taking the average along the string we have   
  \be\label{kv}
   \langle \hat{\bf u}\cdot \dot{\bf y} \rangle = \left\langle A
   |\dot{\bf y}_c| - B |\dot{\bf y}_p| \right\rangle = \left\langle A
   |\dot{\bf y}_c| \left(1-\frac{B}{A}\frac{|\dot{\bf y}_p|}
   {|\dot{\bf y}_c|}\right)\sqrt{\dot{\bf y}^2}\left({\dot{\bf y}_c}^2
   +{\dot{\bf y}_p}^2\right)^{-1/2} \right\rangle = \left\langle  
   \sqrt{\dot{\bf y}^2} \: \frac{ A \left(1-\frac{B}{A}\frac{|\dot
   {\bf y}_p|}{|\dot{\bf y}_c|} \right) } { \left(1+\frac{{\dot{\bf y}_p}^2}
   {{\dot{\bf y}_c}^2} \right)^{1/2} } \right\rangle \,.    
  \ee 
  
  Taking the modulus of (\ref{uhat}) gives $1=A \left(1+\frac{B^2}{A^2}+
  \sum_{i=1}^{D-3} \frac{{C_i}^2}{A^2} \right)^{1/2}$, which we use to 
  substitute for $A$ in (\ref{kv}). Remembering that $kv\simeq \langle 
  \hat{\bf u}\cdot \dot{\bf y} \rangle$ we find 
  \be\label{appk}
   k\simeq \frac{ 1-\frac{B}{A}\frac{v_p}{v_c} } { \left(1+\frac{{v_p}^2}
   {{v_c}^2}\right)^{1/2}\left(1+\frac{B^2}{A^2}+\sum_{i=1}^{D-3}
   \frac{{C_i}^2}{A^2}\right)^{1/2} }  \,.  
  \ee   
 
  For small velocities $(v\ll 1)$ we have that $v\simeq v_c$ and the ratio
  $\frac{v_p}{v_c}$ is much less than unity. However, as the velocity tends 
  to relativistic values, the $v_p$ contribution becomes more and more
  significant. Assuming that this relative contribution is proportional
  to some power of the total velocity, we set $\frac{v_p}{v_c}=fv^{\alpha}$
  with $f$, $\alpha$ constants (clearly $\alpha > 1$). Similarly in the low
  velocity limit, the curvature vector $\bf u$ is parallel to $\dot{\bf y}_c$  
  so $\frac{B}{A},\frac{C_i}{A}\ll 1$ but as $v$ increases, $B$ and $C_i$ 
  become comparable to $A$. Assuming a power law dependence we set $\frac
  {B}{A}=\frac{C_i}{A}=gv^{\beta}$. Equation (\ref{appk}) becomes
  \be\label{appkofv}
   k\simeq \frac{1-fgv^{\alpha+\beta}}{\left(1+f^2 v^{2\alpha}\right)^{1/2}
   \left(1+(D-2)g^2v^{2\beta}\right)^{1/2}} \,.   
  \ee           
  
  Comparison with the helicoidal string solution in D=3 flat space gives 
  $\alpha +\beta=3$ \cite{vosk}. The limit $k(1/\sqrt{2})=0$ (which can be 
  shown to hold analytically \cite{vos}, \cite{thesis}) gives $fg=8$. Then,  
  comparison with the three-dimensional case (\ref{kans3d}) suggests $f\!=\!g$,  
  $\alpha\!=\!\beta$. The approximate formula for $k$ is therefore
  \be\label{appkofvfin}
   k\simeq \frac{1-8v^6}{\left(1+8 v^6\right)^{1/2}
   \left(1+8(D-2)v^6\right)^{1/2}}\,. 
  \ee
        
  For $k_x$ we form $\langle \hat{\bf u}\cdot \dot{\bf x} \rangle$ and
  work as above to find
  \be\label{appk_x}   
  k_x\simeq \frac{v_{xc}}{v_c} \frac{ 1-\frac{B}{A}\frac{{v_{xp}}^2}
  {{v_{xc}}^2}\frac{v_c}{v_p} } { \left( 1+\frac{{v_{xp}}^2}
  {{v_{xc}}^2}\right )^{1/2} \left( 1+\frac{B^2}{A^2}+\sum_{i=1}^{D-3}
  \frac{{C_i}^2}{A^2} \right)^{1/2} }
  \ee
  where we have used that $\langle \hat{\dot{\bf y}}_c\cdot \dot{\bf x}
  \rangle = \left\langle \frac{1}{|\dot{\bf y}_c|} \dot{\bf y}_c\cdot  
  \dot{\bf x} \right\rangle = \left\langle \frac{1}{|\dot{\bf y}_c|}  
  \dot{\bf x}_c \cdot \dot{\bf x} \right\rangle = \left\langle  
  \frac{|{\dot{\bf x}_c}|^2}{|\dot{\bf y}_c|} \right\rangle$.  

  Assuming that $\frac{v_{xp}}{v_{xc}}$ has the same velocity dependence  
  as $\frac{v_p}{v_c}$ we obtain the following approximate formula for 
  $k_x$
  \be\label{appk_xfin}
   k_x\simeq \frac{v_{xc}}{v_c} \frac{1-8v^6}{\left(1+8 v^6\right)^{1/2}
   \left(1+8(D-2)v^6\right)^{1/2}}\,.     
  \ee   
 
  The corresponding equation for $k_\ell$ is
  \be\label{appk_lfin} 
   k_\ell \simeq \frac{v_{\ell c}}{v_c} \frac{1-8v^6}{\left(1+8 v^6 
   \right)^{1/2} \left(1+8(D-2)v^6\right)^{1/2}}\,.     
  \ee   
  
  Equations (\ref{appkofvfin}), (\ref{appk_xfin}) and (\ref{appk_lfin})   
  are indeed the approximate formulae (\ref{kans}-\ref{k_lans}) given  
  in section \ref{evolution}.

\begin{acknowledgments}
We are particularly grateful to Carlos Martins for many illuminating   
discussions about the VOS string network model. We would also like thank    
Fernando Quevedo for fruitful conversations about brane inflation. A.A.   
is also grateful to S.P. Kumar, R.C. Helling and S.A. Hartnoll for    
useful discussions. A.A. is supported by EPSRC, the Cambridge European   
Trust and DAMTP. This work is also supported by PPARC rolling grant   
PPA/G/O/2001/00476.   
\end{acknowledgments}

\bibliography{cstring}

\end{document}